\documentclass[prb,twocolumn,superscriptaddress,showpacs]{revtex4-1}
\usepackage{graphicx, float, hyperref, multirow, verbatim, fontenc, tabularx}

\begin{document}

\title{Thermodynamic stability of alkali metal/zinc double-cation borohydrides at low temperatures}
\author{Tran Doan Huan}
\affiliation{Department of Physics, Universit\"{a}t Basel, Klingelbergstrasse 82, 4056 Basel, Switzerland}
\author{Maximilian Amsler}
\affiliation{Department of Physics, Universit\"{a}t Basel, Klingelbergstrasse 82, 4056 Basel, Switzerland}
\author{Riccardo Sabatini}
\affiliation{Theory and Simulation of Materials, \'{E}cole Polytechnique F\'{e}d\'{e}rale de Lausanne, Station 12, 1015 Lausanne, Switzerland}
\author{Vu Ngoc Tuoc}
\affiliation{Institute of Engineering Physics, Hanoi University of Science and Technology, 1 Dai Co Viet Road, Hanoi, Vietnam}
\author{Nam Ba Le}
\affiliation{Institute of Engineering Physics, Hanoi University of Science and Technology, 1 Dai Co Viet Road, Hanoi, Vietnam}
\affiliation{Department of Physics, University of South Florida, 4202 E. Fowler Ave., Tampa, FL 33620, USA}
\author{Lilia M. Woods}
\affiliation{Department of Physics, University of South Florida, 4202 E. Fowler Ave., Tampa, FL 33620, USA}
\author{Nicola Marzari}
\affiliation{Theory and Simulation of Materials, \'{E}cole Polytechnique F\'{e}d\'{e}rale de Lausanne, Station 12, 1015 Lausanne, Switzerland}
\author{Stefan Goedecker}
\email{stefan.goedecker@unibas.ch}
\affiliation{Department of Physics, Universit\"{a}t Basel, Klingelbergstrasse 82, 4056 Basel, Switzerland}

\date{\today}

\begin{abstract}
We study the thermodynamic stability at low temperatures of a series of alkali metal/zinc double-cation borohydrides, including LiZn(BH$_4$)$_3$,
LiZn$_2$(BH$_4$)$_5$, NaZn(BH$_4$)$_3$, NaZn$_2$(BH$_4$)$_5$, KZn(BH$_4$)$_3$, and KZn$_2$(BH$_4$)$_5$. While LiZn$_2$(BH$_4$)$_5$, NaZn(BH$_4$)$_3$, NaZn$_2$(BH$_4$)$_5$ and KZn(BH$_4$)$_3$ were recently synthesized, LiZn(BH$_4$)$_3$ and KZn$_2$(BH$_4$)$_5$ are hypothetical compounds. Using the minima-hopping method, we discover two new lowest-energy structures for NaZn(BH$_4$)$_3$ and KZn$_2$(BH$_4$)$_5$ which belong to the $C2/c$ and $P2$ space groups, respectively. These structures are predicted to be both thermodynamically stable and dynamically stable, implying that their existence may be possible. On the other hand, we predict that the lowest-energy $P1$ structure of LiZn(BH$_4$)$_3$ is unstable, suggesting a possible reason elucidating why this compound has not been experimentally identified. In exploring the low-energy structures of these compounds, we find that their energetic ordering is sensitive to the inclusion of the van der Waals interactions. We also find that a proper treatment of these interactions, e.g., as given by a non-local density functional such as vdW-DF2, is necessary to address the stability of the low-energy structures of these
compounds.
\end{abstract}

\pacs{61.66.-f, 63.20.dk, 61.05.cp}

\maketitle

\section{Introduction}
Hydrogen is a promising alternative to fossil fuels as it can provide a high density of clean energy. To use hydrogen as a fuel, a safe and efficient storage method is crucial, especially for mobile applications. Many solid state materials, including complex borohydrides $M$(BH$_4$)$_m$, have been studied for the hydrogen-storage purpose (here $M$ is a metal of valence $m$).\cite{schlapbach, mandal, Sakintuna_review, Jain_review,HirscherBook} However, none of these compounds currently meets all of the requirements for a hydrogen-storage system, including the operation temperature of less than $100^\circ$C and the minimum hydrogen charging/discharging rate of 1.2 kg per minute. \cite{DOE_Target} In particular, the kinetics of alkaline-earth metal borohydrides $M$(BH$_4$)$_2$ is generally too slow, taking hours to fully release the hydrogen density stored. \cite{SlowKinetic} Alkali metal borohydrides $M$BH$_4$, on the other hand, are thermodynamically too stable, e.g., LiBH$_4$ starts releasing hydrogen at about $400^\circ$C. \cite{HirscherBook} Attempts to improve their performances, e.g., by using additives, by partial substitutions, or by subjecting them to confinement at the nanoscale, have been met by limited success. \cite{Rude_review}

A large number of double-cation borohydrides $M_iN_j$(BH$_4$)$_{mi+nj}$ were recently synthesized and proposed as candidates for hydrogen-storage materials ($N$ is also a metal of valence $n$). Several examples of such the compounds include Li/K borohydrides, \cite{Nickels:LiK} Li/Ca borohydrides, \cite{Fang:LiCa} Li/Sc borohydrides, \cite{Hagemann:LiSc} Na/Sc borohydrides, \cite{Cerny:NaSc} K/Sc borohydrides, \cite{Cerny:KSc} Na/Al borohydrides, \cite{Lindemann:NaAl} K/Mn and K/Mg borohydrides, \cite{Schouwink_KMnKMg} K/Y borohydrides, \cite{Jaron:KY} Li/Zn and Na/Zn borohydrides, \cite{RavnsbaekLiNa_Zn, Cerny10:LiNa, Ravnsbaek:LiZn} and K/Zn borohydrides. \cite{Cerny12_KZn} The hydrogen capacities of these compounds (8-14\% wt) are comparable with that of Li$_4$BN$_3$H$_{10}$ (11.1\% wt), a hydrogen-rich quaternary compound, \cite{Pinkerton_Li4BN3H10, Herbst_Li4BN3H10} and are relatively higher than the DOE target (5.5\% wt).\cite{DOE_Target} In addition, these double-cation borohydrides were suggested to have decomposition temperatures lying somewhere between those of the constituent single-cation borohydrides. \cite{Jain_review, RavnsbaekLiNa_Zn, Ravnsbaek:LiZn, Cerny12_KZn, Nickels:LiK, Cerny:NaSc, Schouwink_KMnKMg} Such a compelling feature would allow for the adjustment of the decomposition temperature by selecting an appropriate cation combination. Theoretical studies on these newly developed crystalline compounds were subsequently carried out. \cite{Aidhy:LiNaZn, Xiao:LiK, Huang:NaSc, Schouwink_KMnKMg, Kim13:NaSc, Kim12:NaSc}

Of our particular interest here is a series of six Li/Zn, Na/Zn, and K/Zn double-cation borohydrides. Four compounds in this series, e.g., LiZn$_2$(BH$_4$)$_5$, NaZn(BH$_4$)$_3$, NaZn$_2$(BH$_4$)$_5$ and KZn(BH$_4$)$_3$ were experimentally synthesized and their low-energy structures were resolved. \cite{RavnsbaekLiNa_Zn, Cerny10:LiNa, Cerny12_KZn, Ravnsbaek:LiZn} At ambient conditions, LiZn$_2$(BH$_4$)$_5$ crystallizes in an orthorhombic $Cmca$ (no. 64) phase as first determined by powder x-ray diffraction (XRD) analysis \cite{RavnsbaekLiNa_Zn} and then refined by powder neutron diffraction, Raman and NMR spectroscopy. \cite{Ravnsbaek:LiZn, Cerny10:LiNa} NaZn(BH$_4$)$_3$ and NaZn$_2$(BH$_4$)$_5$ were found to be in two phases which both belong to the monoclinic $P2_1/c$ space group (no. 14). \cite{RavnsbaekLiNa_Zn, Cerny10:LiNa} At 100K, KZn(BH$_4$)$_3$ was determined to crystallize in a rhombohedral $R3$ phase (no. 146). \cite{Cerny12_KZn} The other two borohydrides of the series, e.g., LiZn(BH$_4$)$_3$ and KZn$_2$(BH$_4$)$_5$, are hypothetical compounds. While both of them have not experimentally been observed yet, a $P2_1/c$ phase (no. 14) and a $P1$ (no. 1) phase of LiZn(BH$_4$)$_3$ were theoretically proposed in Refs. \onlinecite{Cerny10:LiNa} and \onlinecite{Aidhy:LiNaZn}, respectively. No information on KZn$_2$(BH$_4$)$_5$ is available in the literature.

Some of the structural phases experimentally resolved for these compounds were recently predicted to be thermodynamically unstable at 0K by {\it ab initio} calculations in which the vibrational energies were neglected. \cite{Aidhy:LiNaZn} In particular, the unrefined $Cmca$ phase of LiZn$_2$(BH$_4$)$_5$ proposed by Ref. \onlinecite{RavnsbaekLiNa_Zn} is unstable with respect to the decomposition into Zn(BH$_4$)$_2$ and the $P1$ phase of LiZn(BH$_4$)$_3$. Furthermore, the $P2_1/c$ phase of NaZn(BH$_4$)$_3$ is unstable with respect to the decomposition into NaBH$_4$ + Zn(BH$_4$)$_2$. However, because the refined $Cmca$ structure of LiZn$_2$(BH$_4$)$_5$ was then noted \cite{Aidhy:LiNaZn} to be lower in energy than the unrefined $Cmca$ structure which was used to predict the stability of the Li/Zn borohydrides, two of the theoretical predictions of Ref. \onlinecite{Aidhy:LiNaZn} related to these compounds have to be revisited. The first one is that LiZn$_2$(BH$_4$)$_5$ is unstable, and the second one is that the $P1$ structure of the hypothetical borohydride LiZn(BH$_4$)$_3$ is stable. In addition, the recent proposal of the $P1$ phase of NaZn(BH$_4$)$_3$, which was theoretically predicted \cite{Aidhy:LiNaZn} to be thermodynamically more stable than the experimentally synthesized $P2_1/c$ phase, \cite{RavnsbaekLiNa_Zn} suggests that unexplored low-energy structures of the Li/Zn and Na/Zn borohydrides may exist. Finally, as KZn(BH$_4$)$_3$ was synthesized very recently, \cite{Cerny12_KZn} it is worth exploring the (additional) possible low-energy structures of KZn(BH$_4$)$_3$ and its hypothetical related compounds, e.g., KZn$_2$(BH$_4$)$_5$.

In this work, we report a first-principles study on the thermodynamic stability of the series of Li/Zn, Na/Zn, and K/Zn double-cation borohydrides. By using the experimental data to validate the calculated results, we find that van der Waals (vdW) interactions have to be properly treated using the vdW density-functional theory. \cite{Dion-DF, Thonhauser-DF, Roman-DF, vdW-DF2} To discuss the stability of these compounds, we use the minima-hopping method \cite{Goedecker:MHM, Amsler:MHM} to extensively explore the low-energy structures of these compounds. By determining the vibrational free energy from the phonon frequency spectra obtained for the examined structures, we study the thermodynamic stability of these compounds at finite temperatures.

\section{Methods}

First-principles calculations reported in this work were performed with \textit{Vienna Ab Initio Simulation Package} ({\sc vasp}) \cite{vasp1,vasp2,vasp3,vasp4} at the density functional theory (DFT) \cite{dft1, dft2} level, employing the projected augmented wave formalism. \cite{paw} The valence electron configurations of Zn, B, Li, Na, and K used for our calculations are $3d^{10} 4s^2$, $2s^2 2p^1$, $1s^2 2s^1$, $2s^2 2p^6 3s^1$ and $3s^2 3p^6 4s^1$, respectively. The convergence of the DFT total energies was ensured by a Monkhorst-Pack $\bf k$-point mesh \cite{monkhorst} from $5 \times 5 \times 5$ to $9 \times 9 \times 9$, depending on the simulation cell sizes, for sampling the Brillouin zone and a kinetic energy plane wave cutoff of 900 eV. Atomic and cell variables were simultaneously relaxed until the residual forces were smaller than $0.01$ eV/\AA. The space groups of the examined structures were determined by {\sc findsym} \cite{findsym} while some figures were prepared by {\sc vesta}.\cite{vesta}

To test our computational scheme, we optimized the experimentally reported structures of LiZn$_2$(BH$_4$)$_5$, \cite{RavnsbaekLiNa_Zn, Ravnsbaek:LiZn, Cerny10:LiNa} NaZn(BH$_4$)$_3$, \cite{RavnsbaekLiNa_Zn, Cerny10:LiNa} NaZn$_2$(BH$_4$)$_5$, \cite{RavnsbaekLiNa_Zn, Cerny10:LiNa} KZn(BH$_4$)$_3$, \cite{Cerny12_KZn} LiBH$_4$, \cite{soulieLi} NaBH$_4$, \cite{FischerNa} and KBH$_4$. \cite{Renaudin_KBH4, VajeestonNaK} Three conventional
exchange-correlation functionals, i.e., Perdew-Burke-Ernzerhof (PBE) generalized gradient approximation (GGA), \cite{PBE} PBEsol GGA, \cite{PBEsol} and local density approximation (LDA), were used. Because the vdW interactions were recently pointed out \cite{Bill:vdW} to play an important role in determining the stability of magnesium borohydride Mg(BH$_4$)$_2$ at low energies, we also tested our computational scheme with three vdW implementations in {\sc vasp}, i.e., the DFT-D2 method of Grimme, \cite{DFT-D2} the DFT-TS method of Tkatchenko-Scheffler, \cite{DFT-TS} and the modified Langreth-Lundqvist non-local density functional (vdW-DF2). \cite{Dion-DF, Thonhauser-DF, Roman-DF, vdW-DF2} Besides the lattice parameters, we examined the relative volume change $\Delta V \equiv \left[V^{\rm (t)}-V^{\rm (e)}\right]/V^{\rm (e)}$ where $V^{\rm (t)}$ and $V^{\rm (e)}$ are the volumes of the theoretically and experimentally determined unit cells. With GGA, $\Delta V$ follows a $t$ location-scale distribution with the median at $\Delta V= 3.2\%$, indicating the tendency of GGA to overestimate the unit cell volume. \cite{volume_change}

The optimized structures are summarized in the Supplemental Material \cite{supplement} and in Table \ref{table_validate}. With PBE, the volume changes $\Delta V$ obtained for LiBH$_4$, NaBH$_4$, KBH$_4$, and LiZn$_2$(BH$_4$)$_5$ are reasonable. For the other compounds, $V^{\rm (t)}$ are considerably larger than $V^{\rm (e)}$ so that $\Delta V$ is as high as $16\%$. Being consistent with the results reported in Ref. \onlinecite{Aidhy:LiNaZn}, the lattice parameter $b$ calculated for the $P2_1/c$ structure of NaZn(BH$_4$)$_3$ is about $15\%$ larger than the experimental data. In addition, the lattice parameters $a$ of NaZn$_2$(BH$_4$)$_5$ and $c$ of KZn(BH$_4$)$_3$ are overestimated by about $10\%$. On the other hand, $V^{\rm (t)}$ are significantly smaller than $V^{\rm (e)}$ for all of the compounds when LDA is used. Two methods for evaluating the vdW interactions, i.e., DFT-D2 and DFT-TS, also consistently underestimate the lattice parameters, resulting in considerably large values of $\Delta V$, which can exceed $-20\%$ in some cases.

\begin{table*}[t]
\begin{center}
\caption{Geometrical parameters (lattice parameters $a$, $b$, and $c$ in Angstrom, angle $\beta$ in degree, and volume difference $\Delta V$ in percents) obtained by computationally optimizing the experimentally-reported low-energy structures of the examined compounds and several related compounds. References are given for the experimental data, which are presented for the comparison purpose. } \label{table_validate}
\begin{tabular}{llrrrrrrrrrrrrrrrrr}
\hline
\hline
\multirow{2}{*}{Compound}&\multirow{2}{*}{Space group}&\multicolumn{5}{c}{Calculations with PBE}&&\multicolumn{5}{c}{Calculations with vdW-DF2} & & \multicolumn{5}{c}{Experiments}\\
\cline{3-7}\cline{9-13}\cline{15-19}
&& $a$& $b$ & $c$& $\beta$ &$\Delta V$& &$a$ & $b$&$c$&$\beta$ & $\Delta V$&  &$a$&$b$&$c$&$\beta$& Ref.\\
\hline
LiZn$_2$(BH$_4$)$_5$ &  $Cmca$  & 8.63 & 18.03 & 15.43 & 90 & 2.0 & & 8.71 & 17.50 & 15.67 & 90 & 0.5 & & 8.59 & 17.86 & 15.35 & 90 &
  [\onlinecite{Ravnsbaek:LiZn}]\\
NaZn(BH$_4$)$_3$ & $P2_1/c$ & 8.02  &  5.38   & 18.57 & 101.5 &14.3& & 8.83 & 4.62 & 16.89 & 99.8& $-1.2$ & & 8.27 & 4.52 & 18.76 & 101.7 &
  [\onlinecite{RavnsbaekLiNa_Zn}]\\
NaZn$_2$(BH$_4$)$_5$  & $P2_1/c$  & 10.45 & 16.19 & 9.06 & 112.4& 6.4 & & 8.69 & 17.12 & 9.34 &111.5&$-1.7$& & 9.40 & 16.64 & 9.14 & 112.7 &
  [\onlinecite{RavnsbaekLiNa_Zn}]\\
KZn(BH$_4$)$_3$ &  $R3$ & 7.80 & 7.80  & 12.24 & 90  &16.5&&7.65&7.65&11.80&90&8.1&& 7.63  & 7.63  & 10.98 & 90 &
  [\onlinecite{Cerny12_KZn}]\\
LiBH$_4$ &   $Pnma$ & 7.30 &4.39 &6.60 & 90 & $-2.4$& & 6.90 & 4.50 & 6.72 & 90 &$-3.7$ & & 7.18  & 4.44  & 6.80 & 90 &
  [\onlinecite{soulieLi}]\\
NaBH$_4$ &   $P4_2/nmc$    & 4.34 & 4.34 & 5.88 & 90 & 0.6 & & 4.35 & 4.35 & 5.85 & 90 & 0.6 & & 4.33 & 4.33 & 5.87 & 90 &
  [\onlinecite{FischerNa}]\\
KBH$_4$ &   $P4_2/nmc$  & 4.76  & 4.76 & 6.68 & 90 & 3.8 & & 4.72 & 4.72 & 6.60 &90 & 0.9 & & 4.70 & 4.70 & 6.60 & 90 &
  [\onlinecite{Renaudin_KBH4}]\\
\hline
\hline
\end{tabular}
\end{center}
\end{table*}

More reasonable results for $V^{\rm (t)}$ were obtained from the calculations using PBEsol and vdW-DF2. With PBEsol, $\Delta V$ obtained for the $P2_1/c$ structures of NaZn(BH$_4$)$_3$ and NaZn$_2$(BH$_4$)$_5$ are small while the unit cell of the $R3$ structure of KZn(BH$_4$)$_3$ is expanded with a considerable volume change $\Delta V=5.7\%$. For other compounds, PBEsol underestimates the unit cell volumes by up to $7.6\%$. The agreement between $V^{\rm (e)}$ and $V^{\rm (t)}$ optimized with vdW-DF2 is generally better. A detailed comparison between the examined structures optimized by PBEsol and vdW-DF2 is given in Table II of the Supplemental Material. \cite{supplement} Although $\Delta V$ obtained with PBEsol for the experimental $P2_1/c$ structure of NaZn$_2$(BH$_4$)$_5$ is small ($-0.7\%$), the optimized unit cell is strongly distorted. Compared to the experimental data, $a$ and $c$ are underestimated by $13.8\%$ and $5.9\%$, respectively, while $b$ is overestimated by $23.5\%$. Smaller distortions can also be observed in several other structures optimized with PBEsol. Nevertheless, with vdW-DF2, the distortions of the optimized structures are generally smaller. Table \ref{table_validate} also shows that the results obtained with vdW-DF2 are much closer to the experimental data than those obtained with PBE.

To confirm the role of vdW-DF2 in bringing the optimized structures closer to the experimentally measured data, we have optimized the experimental structures examined in Table \ref{table_validate} by using the PWSCF code of the {\sc quantum espresso} distribution. \cite{QE} In the calculations, we used the rPW86 pseudopotentials \cite{PW86, rPW86} of the PsLibrary at an energy cutoff of 1100 eV and the vdW-DF2 density functional for the vdW interactions. The lattice parameters of the optimized structures are shown in Table III of the Supplemental Material, \cite{supplement} clearly demonstrating that the results obtained from calculations by {\sc quantum espresso} and {\sc vasp} are in good agreement. These observations indicate that the long-ranged vdW interactions in the examined compounds are well captured by the non-local vdW-DF2 functional. Therefore, in addition to the DFT energies $E_{\rm DFT}^{\rm PBE}$ at the PBE level, we also calculated $E_{\rm DFT}^{\rm vdW}$, the DFT energies with vdW corrections obtained with vdW-DF2.

It is worth noting that the lattice parameters shown in Table \ref{table_validate} were calculated at 0K and compared with the experimental data measured at various finite temperatures. Therefore, some discrepancies between the theoretical results and the corresponding experimental values observed in Table \ref{table_validate} may partially be related to the thermal expansion of the examined compounds. To calculate the thermal expansion coefficient of a given crystalline compound, one has to go beyond the harmonic approximation assumed in this work. For such purposes, the quasi-harmonic approximation, described elsewhere, \cite{BaroniQHA,Wrobel_2012} is an approach for calculating such a quantity.

Unconstrained searches for low-energy structures of the examined compounds were performed by the minima-hopping method (MHM). \cite{Goedecker:MHM,Amsler:MHM} In this method, the DFT energy landscapes are explored by consecutive short molecular dynamics steps followed by local geometry relaxations. The initial velocities for the molecular dynamics runs are chosen approximately along soft mode directions, allowing efficient escapes from local minima, and aiming towards the global minimum. This method was successfully applied in a wide range of material structure predictions. \cite{hellmann_2007, roy_2009, bao_2009, willand_2010, de11, amsler_crystal_2011, Livas, amsler12, HuanZn, HuanAlanates} It is worth to note that there are compounds, e.g., Li$_2$NH,\cite{Hector_DFT} of which the experimentally synthesized structure is different from that predicted by DFT. Therefore, the low-energy structures predicted by a theoretical method such as MHM need validations by experiments. In fact, some of the theoretical predictions by MHM on neutral Si clusters and four-fold defects in silicon were recently confirmed by experiments. \cite{haertelt, Markevich}

Phonon frequency spectra of the examined structures were calculated with the PBE functional using the {\sc phonopy} package. \cite{phonopy, phonopy_sc} Given a sufficiently large relaxed super cell, finite atomic displacements with an amplitude of 0.01 \AA~ were introduced. The atomic forces within the super cells were calculated with {\sc vasp} and the phonon frequencies were then calculated from the dynamical matrix, given in terms of the force constants. \cite{phonopy_sc} The longitudinal optical/transverse optical splitting was not considered because its effects were reported to be negligible for a wide variety of hydrides. \cite{hector07, herbst10}

For a double-cation borohydride, we computed the formation energy to examine if it is stable with respect to the decomposition into the corresponding single-cation borohydrides and related compounds. As an example, the formation energy $\Delta F_T$ at a given temperature $T$ of LiZn$_2$(BH$_4$)$_5$ was determined by
\begin{equation}\label{Eq:mixing}
\begin{array}{ll}
\Delta F_T & = F_T[{\rm LiZn_2(BH_4)_5}] \\
\\
& - \{(1-x)F_T[{\rm LiBH_4}] + xF_T[{\rm Zn(BH_4)_2}]\}
\end{array}
\end{equation}
where $x=2/3$ is the Zn cation composition of LiZn$_2$(BH$_4$)$_5$. Here $F_T[{\rm LiZn_2(BH_4)_5}]$, $F_T[{\rm LiBH_4}]$, and $F_T[{\rm Zn(BH_4)_2}]$ are the Helmholtz free energies (including the vibrational energies) at the temperature $T$ of one formula unit of LiZn$_2$(BH$_4$)$_5$, LiBH$_4$, and Zn(BH$_4$)$_2$, respectively. The examined structure of LiZn$_2$(BH$_4$)$_5$ is stable if $\Delta F_T$ is negative; otherwise, it would decompose into LiBH$_4$ and Zn(BH$_4$)$_2$. At a finite temperature $T$, the vibrational free energies were determined within the harmonic approximation from the phonon frequency spectra of the examined structures. Using Eq. \ref{Eq:mixing}, we also computed the formation energies $\Delta E_{\rm DFT}^{\rm PBE}$ and $\Delta E_{\rm DFT}^{\rm vdW}$ from the DFT energies $E_{\rm DFT}^{\rm PBE}$ and $E_{\rm DFT}^{\rm vdW}$. To determine the formation energies according to Eq. \ref{Eq:mixing}, we used the $I4_122$ phase (no. 98) of Zn(BH$_4$)$_2$, \cite{Aidhy:LiNaZn, HuanZn, JeonZnBH4, SrinivasanZnBH4} the $Pnma$ phase (no. 62) of LiBH$_4$, \cite{soulieLi} and the $P4_2/mnc$ phase (no. 137) of both NaBH$_4$ and KBH$_4$. \cite{Renaudin_KBH4,FischerNa, VajeestonNaK, zutel07}

\section{Low-temperature structures of alkali metal/zinc double-cation borohydrides}
We searched for low-energy structures of LiZn(BH$_4$)$_3$, NaZn(BH$_4$)$_3$, KZn(BH$_4$)$_3$, LiZn$_2$(BH$_4$)$_5$, NaZn$_2$(BH$_4$)$_5$, and KZn$_2$(BH$_4$)$_5$ by using the minima-hopping method. For the first three compounds, we searched for 1 f.u. (17 atoms) and 2 f.u. (34 atoms) in a cell while for the last three compounds, searches were performed with 1 f.u. (28 atoms) in a cell. Starting from random input structures, we discovered a large number of new low-energy structures for these compounds. Additional searches were also carried out by relaxing several lowest-energy structures predicted for each compound after replacing their cations by the cations of other appropriate species. Our MHM runs were performed with the PBE functional.

\begin{table*}[t]
\begin{center}
\caption{Summary of the available structures of alkali metal/zinc double cation borohydrides. $\Delta E^{\rm PBE}_{\rm DFT}$, $\Delta F^{\rm PBE}_{0K}$, $\Delta F^{\rm PBE}_{\rm 100K}$, and $E_{\rm DFT}^{\rm vdW}$, given in kJ/mol cation, are the energies of formation determined from $E^{\rm PBE}_{\rm DFT}$, $F_{\rm 0K}$, $F_{\rm 100K}$, and $E_{\rm DFT}^{\rm vdW}$, respectively. These structures are indicated by their space groups. Densities $\rho$ are given in $\rm g/cm^3$. References are shown for structures not originating from this work.} \label{table_energy}
\begin{tabular}{llrrrrrr}
\hline
\hline
Compounds& Space group (no.) & $\rho({\rm g/cm^3})$ & $\Delta E^{\rm PBE}_{\rm DFT} $ & $\Delta F^{\rm PBE}_{\rm 0K}$ & $\Delta F^{\rm PBE}_{\rm 100K}$ & $\Delta E_{\rm DFT}^{\rm vdW}$ & References \\
\hline
LiZn(BH$_4$)$_3$     & $P1$ (1)                & 0.80 & $-6.70$  & $-6.89$  & $-7.45$  & $-3.93$  &
   [\onlinecite{Aidhy:LiNaZn}]\\
LiZn$_2$(BH$_4$)$_5$ & $Cmca$ (64)             & 1.18 & $-10.14$ & $-10.67$ & $-10.61$ & $-18.83$ &
   [\onlinecite{Cerny10:LiNa}], [\onlinecite{Ravnsbaek:LiZn}]\\
                     & $I\overline{4}m2$ (119) & 0.57 & $-11.70$ & $-12.96$ & $-13.59$ & $ 0.03 $  &\\
NaZn(BH$_4$)$_3$     & $P2_1/c$ (14)           & 1.28 & $-6.55$  & $-8.43$  &  $-8.74$ & $-5.50$  &
   [\onlinecite{RavnsbaekLiNa_Zn}]\\
                     & $P1$ (1)                & 0.91 & $-7.83$  & $-9.25$  & $-9.92$  & $-4.50$ &
   [\onlinecite{Aidhy:LiNaZn}]\\
                     & $C2/c$ (15)             & 1.00 & $-10.90$ & $-11.76$ & $-12.05$ & $-9.39$  & \\
NaZn$_2$(BH$_4$)$_5$ & $P2_1/c$ (14)           & 1.15 & $-12.08$ & $-13.91$ & $-14.58$ & $-11.50$ &
   [\onlinecite{Ravnsbaek:LiZn}] \\
                     & $I\overline{4}m2$ (119) & 0.54 & $-11.87$ & $-13.58$ & $-14.87$ & 3.16  &\\
KZn(BH$_4$)$_3$      & $R3$ (146)              & 1.15 & $-7.31$  & $-9.28$  & $-10.21$ & $-9.16$ &
   [\onlinecite{Cerny12_KZn}]\\
                     &$P6_3/m$ (176)           & 1.05 & $-11.08$ & $-12.60$ & $-13.32$ & $-4.95$ &\\
KZn$_2$(BH$_4$)$_5$  &$P2$ (3)                 & 1.09 & $-8.84$  & $-10.75$ & $-11.39$ & $-11.05$ &\\
\hline
\hline
\end{tabular}
\end{center}
\end{table*}

\begin{figure}[b]
  \begin{center}
    \includegraphics[width=8.25cm]{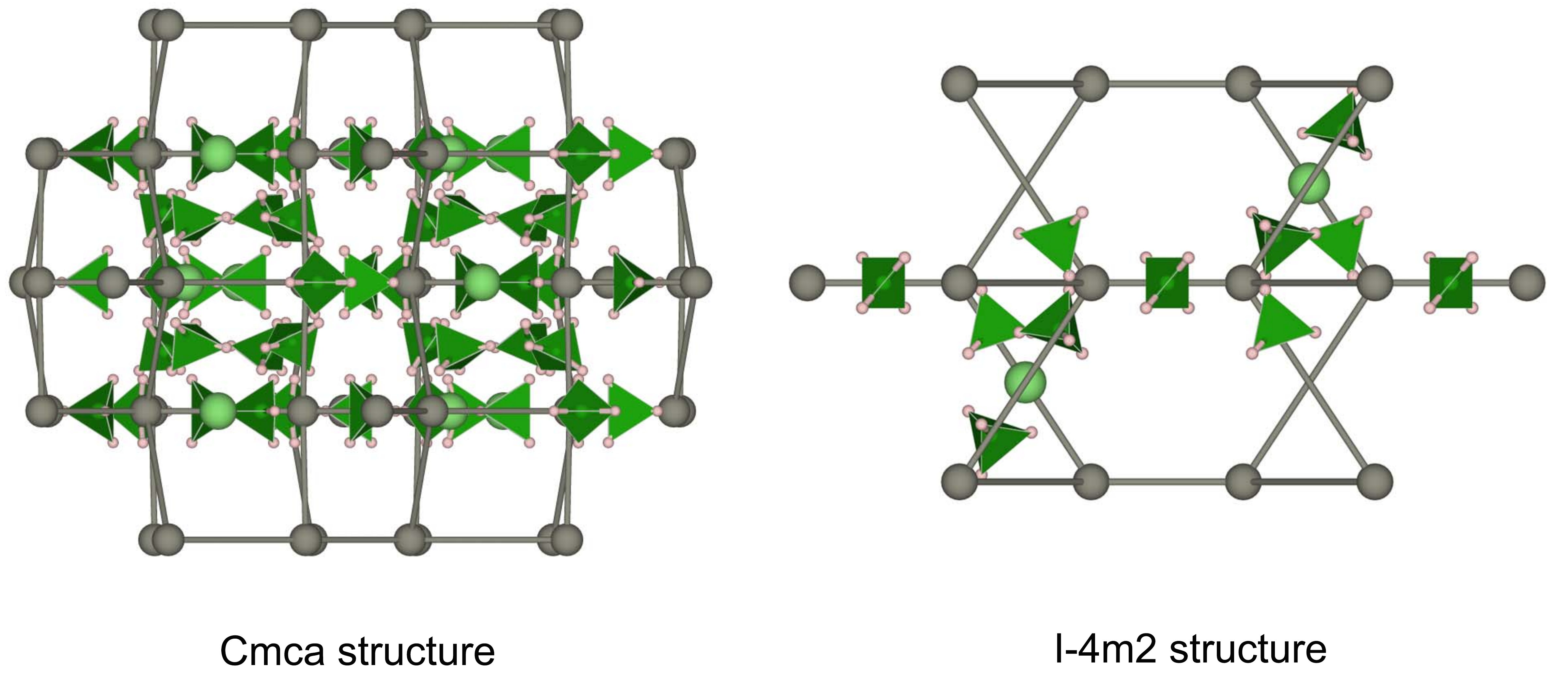}
  \caption{(Color online) $Cmca$ (left) and $I\overline{4}m2$ (right) phases of LiZn$_2$(BH$_4$)$_5$. Lithium and zinc atoms are shown by green and gray spheres while [BH$_4$]$^-$ complex anions are shown by green tetrahedra. Zn-Zn interlinked ``bonds" are used as guides for the eyes.}\label{fig:LiZn2}
  \end{center}
\end{figure}

\begin{figure}[b]
  \begin{center}
    \includegraphics[width=8cm]{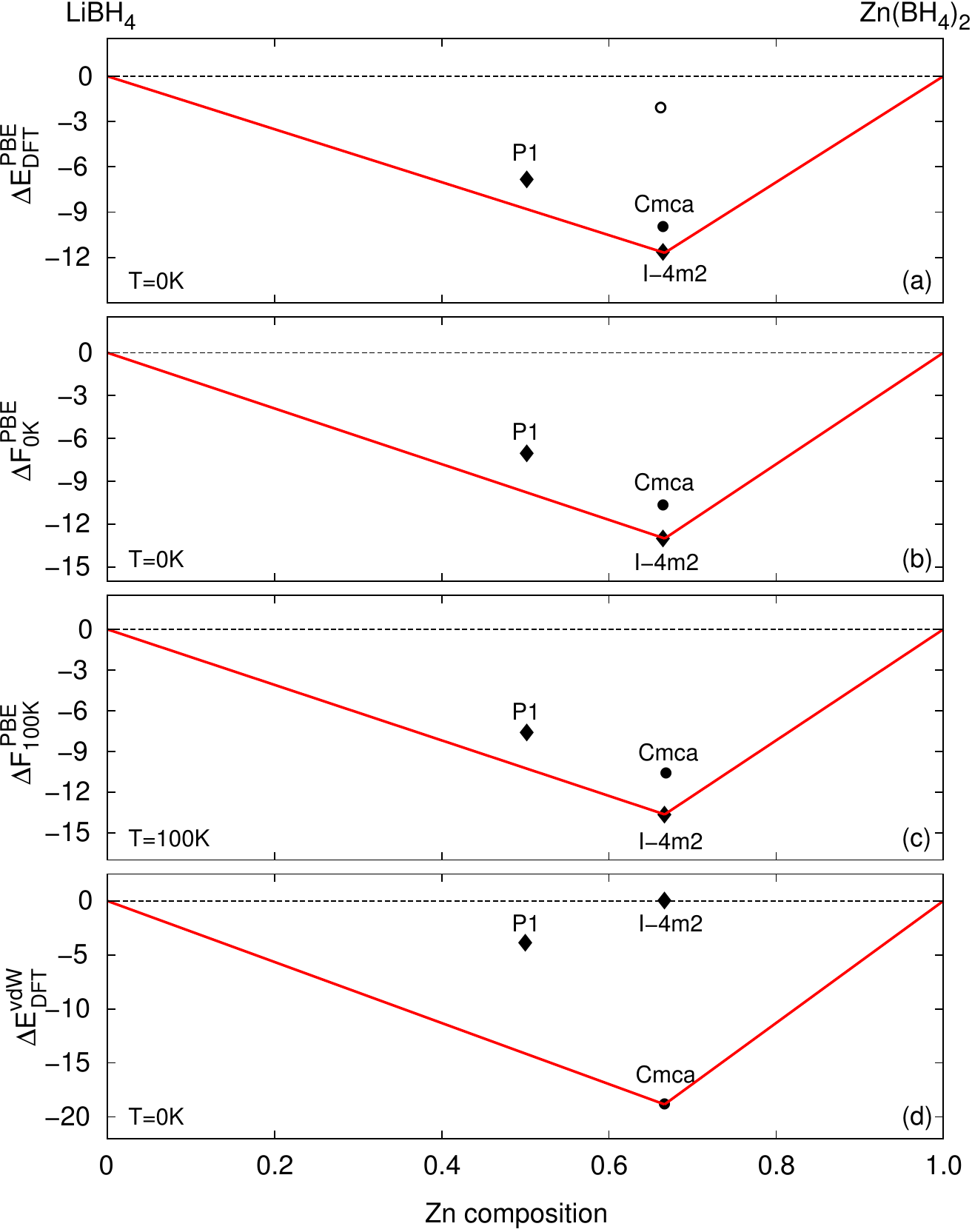}
  \caption{(Color online) Formation energies $\Delta E^{\rm PBE}_{\rm DFT}$,   $\Delta F^{\rm PBE}_{\rm 0K}$, $\Delta F^{\rm PBE}_{\rm 100K}$, and $\Delta E^{\rm vdW}_{\rm DFT}$ (in units of kJ/mol cation) of Li/Zn double-cation borohydrides shown vs. $x$, the Zn composition of the compounds. ForLiZn(BH$_4$)$_3$, $x=1/2$  while for LiZn$_2$(BH$_4$)$_5$, $x=2/3$. The structures studied are indicated by symbols with the space groups nearby. Diamonds/circles are used for theoretically-predicted/experimentally-reported phases. Open circle represents the \emph{unrefined} $Cmca$ structure of LiZn$_2$(BH$_4$)$_5$. Constructed from the thermodynamically most stable structures, the red-solid lines form a convex hull, above which all structures are unstable. } \label{fig:stable_LiZn}
  \end{center}
\end{figure}

We show in Table \ref{table_energy} the densities and the formation energies, determined with PBE and vdW-DF2, of all the lowest-energy structures predicted together with those already reported in the literature. We note that our results of $\Delta E^{\rm PBE}_{\rm DFT}$ for the $P1$ phase of LiZn(BH$_4$)$_3$ and the for the unrefined $Cmca$ phase of LiZn$_2$(BH$_4$)$_5$ are consistent with those reported by Ref. \onlinecite{Aidhy:LiNaZn}. The crystallographic information of all the structures reported in this work can be found in the Supplemental Material. \cite{supplement} Simulated XRD patterns performed by {\sc fullprof} \cite{fullprof} and phonon densities of states calculated for all the new structures are also shown in the Supplemental Material, indicating that they are dynamically stable and are different from those which can be located in the  literature.

\subsection{Lithium/zinc borohydrides}
We optimized the refined $Cmca$ structure of LiZn$_2$(BH$_4$)$_5$ and found that this structure is thermodynamically more stable than the unrefined $Cmca$ structure by $\simeq 7.8$ kJ/mol cation. Additionally, from the MHM searches for LiZn$_2$(BH$_4$)$_5$, we discovered a new tetragonal $I\overline{4}m2$ structure (no. 119), which is more stable than the refined $Cmca$ structure by $\simeq 1.5$ kJ/mol cation. Further calculations with the PBE functional indicated that $I\overline{4}m2$ is the thermodynamically most stable structure of LiZn$_2$(BH$_4$)$_5$ at temperatures up to 100K (see Table \ref{table_energy}). While the $Cmca$ structure consists of two identical inter-penetrated three-dimensional (3D) frameworks, \cite{RavnsbaekLiNa_Zn, Cerny10:LiNa, Ravnsbaek:LiZn} the $I\overline{4}m2$ structure consists of one of them with differences in the sequence of Zn atoms and Li atoms. Therefore, it is not surprising that the $I\overline{4}m2$ structure has a very low density with $\rho = 0.57~{\rm g \times cm}^{-3}$, about one half of the density $\rho = 1.18~ {\rm g \times cm}^{-3}$ of the $Cmca$ structure. An illustration for the $Cmca$ and $I\overline{4}m2$ structures is given in Fig. \ref{fig:LiZn2}, clearly showing the differences in geometry and in density between them.

For the hypothetical compound LiZn(BH$_4$)$_3$, we optimized the $P2_1/c$ and $P1$ structures which were theoretically proposed in Refs. \onlinecite{Cerny10:LiNa} and \onlinecite{Aidhy:LiNaZn}, respectively. We found that for the $P1$ structure, $\Delta E^{\rm PBE}_{\rm DFT} \simeq -6.7$ kJ/mol cation while for the $P2_1/c$ structure, $\Delta E^{\rm PBE}_{\rm DFT} \simeq 19 $ kJ/mol cation. The positive formation energy of the $P2_1/c$ structure hints that this structure is unstable and is an incorrect structural model. \cite{Cerny_commun_mixed} On the other hand, $P1$ is the lowest-energy structure of LiZn(BH$_4$)$_3$, as also confirmed by the MHM runs we performed for this hypothetical compound.

We then examined the thermodynamic stability of the $P1$ structure of LiZn(BH$_4$)$_3$ and the $Cmca$ and the $I\overline{4}m2$ structures of LiZn$_2$(BH$_4$)$_5$ based on the convex hull constructed from the formation energies of the thermodynamically most stable structures calculated at 0K and 100K with the PBE functional. Fig. \ref{fig:stable_LiZn}(a) shows that in agreement with the prediction of Ref. \onlinecite{Aidhy:LiNaZn}, the unrefined $Cmca$ phase is unstable with respect to the decomposition into LiZn(BH$_4$)$_3$ and Zn(BH$_4$)$_2$. Because the refined $Cmca$ and $I\overline{4}m2$ structures of LiZn$_2$(BH$_4$)$_5$ are much lower in energy than the unrefined $Cmca$ phase, they remain thermodynamically stable at temperatures up to 100K [see Figs. \ref{fig:stable_LiZn}(a)-(c)]. The $P1$ structure of the hypothetical compound LiZn(BH$_4$)$_3$, on the other hand, is always unstable  and would decompose  into LiBH$_4$ and LiZn$_2$(BH$_4$)$_5$ within this temperature range.

Figs. \ref{fig:stable_LiZn}(a)-(c) demonstrate that at temperatures up to 100K, lattice vibrations have essentially no effect on the energetic ordering. Therefore, it is expected that the formation energy $\Delta E_{\rm DFT}^{\rm vdW}$ determined from $E_{\rm DFT}^{\rm vdW}$ can be used to establish more accurately the stability of these structures. Fig. \ref{fig:stable_LiZn}(d) shows that by considering the vdW interactions, the energetic ordering of the $I\overline{4}m2$ and the $Cmca$ phases of LiZn$_2$(BH$_4$)$_5$ was reverted. In particular, the $Cmca$ phase is thermodynamically more stable than the $I\overline{4}m2$ phase by $\simeq 18$kJ/mol cation while the $P1$ phase of LiZn(BH$_4$)$_3$ is, again, unstable and would decompose into LiBH$_4$ and LiZn$_2$(BH$_4$)$_5$. Together with the fact that attempts to experimentally identify LiZn(BH$_4$)$_3$ have failed,  \cite{RavnsbaekLiNa_Zn, Cerny10:LiNa} Fig. \ref{fig:stable_LiZn} may be viewed as a supporting evidence of the hypothesis that LiZn(BH$_4$)$_3$  is actually unstable.

We suggest that the significant role of the vdW interactions in determining the energetic ordering of the $Cmca$ and the $I\overline{4}m2$ structures can be traced back to the difference in the number of 3D frameworks between them. Because there are no bonds between the two 3D frameworks of the $Cmca$ structure \cite{RavnsbaekLiNa_Zn, Cerny10:LiNa, Ravnsbaek:LiZn} while the $I\overline{4}m2$ structure has only one 3D framework, they are almost similar in their energies at the PBE level. However, in the vdW-DF2 calculations, long-ranged interactions between the two frameworks of the $Cmca$ structure can be captured, lowering the energy compared to that of the $I\overline{4}m2$ structure. This behavior is similar to that reported by Ref. \onlinecite{Bill:vdW} where it was argued that the PBE functional artificially stabilizes structures of Mg(BH$_4$)$_2$ with unusually low densities, and a treatment within non-local density functionals such as vdW-DF would be a solution to the problem. In this work, we show that to access the stability of LiZn$_2$(BH$_4$)$_5$ at low temperatures, it is obviously desirable to properly incorporate the vdW interactions by using a non-local density functional such as vdW-DF2.

\subsection{Sodium/zinc borohydrides}
\begin{figure}[t]
  \begin{center}
    \includegraphics[width=8.25 cm]{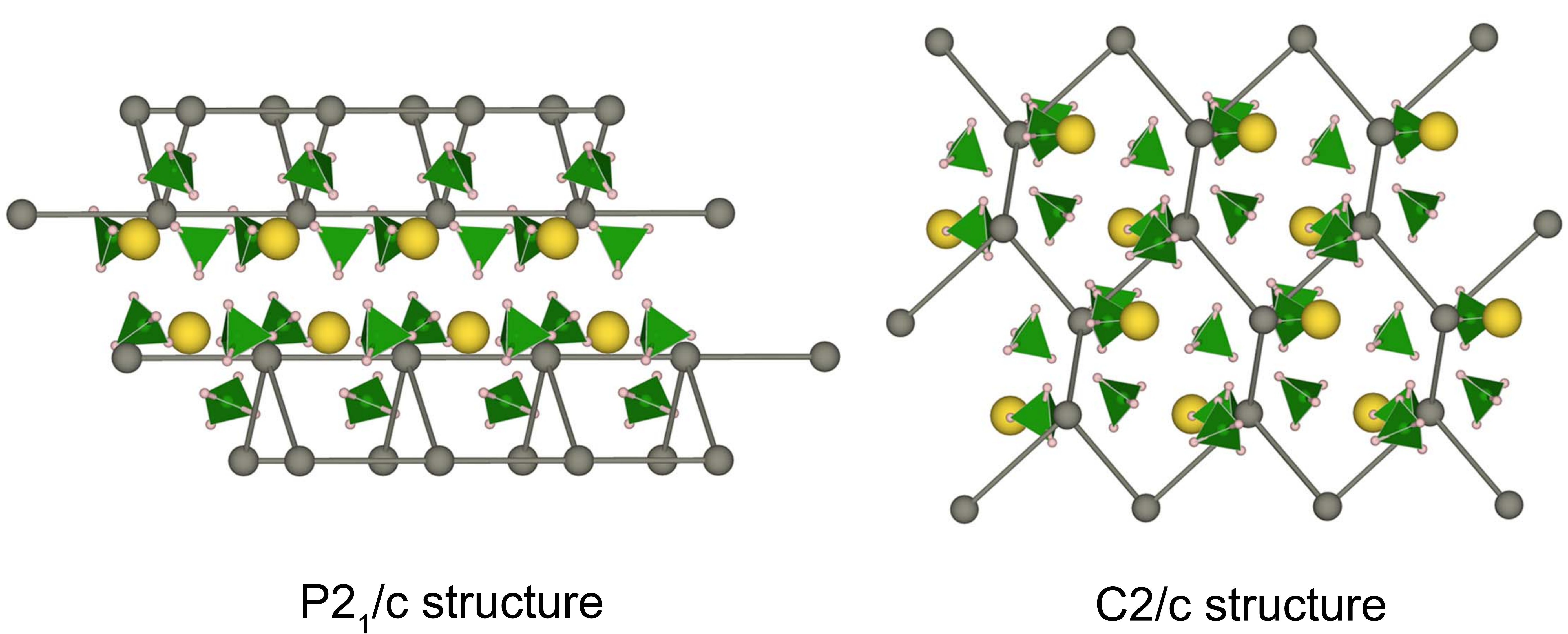}
  \caption{(Color online) Two low-energy structures of NaZn(BH$_4$)$_3$: $P2_1/c$ (left) and $C2/c$ (right). Sodium and zinc and atoms are shown by yellow and gray spheres while [BH$_4$]$^-$ complex anions are shown by green tetrahedra. }\label{fig:NaZn}
  \end{center}
\end{figure}

For NaZn(BH$_4$)$_3$, the experimentally observed $P2_1/c$ structure was theoretically predicted \cite{Aidhy:LiNaZn} to be thermodynamically unstable at 0K. A new $P1$ structure was then proposed \cite{Aidhy:LiNaZn} to be lower in energy than the $P2_1/c$ structure. Using the MHM, we discovered a new monoclinic $C2/c$ structure (no. 15), which is lower than the $P1$ structure by $\simeq$ 3kJ/mol cation. Our further results calculated with the PBE functional indicate that $C2/c$  is the thermodynamically most stable structure at temperatures up to 100 K. The $C2/c$ phase, of which the density is $\rho = 1.00~ {\rm g \times cm}^{-3}$, is slightly lighter than the $P2_1/c$ phase with $\rho = 1.28~ {\rm g \times cm}^{-3}$. As an illustration, the geometries of the $P2_1/c$ and $C2/c$ phases are shown in Figure \ref{fig:NaZn}.

\begin{figure}[b]
  \begin{center}
    \includegraphics[width=8.25 cm]{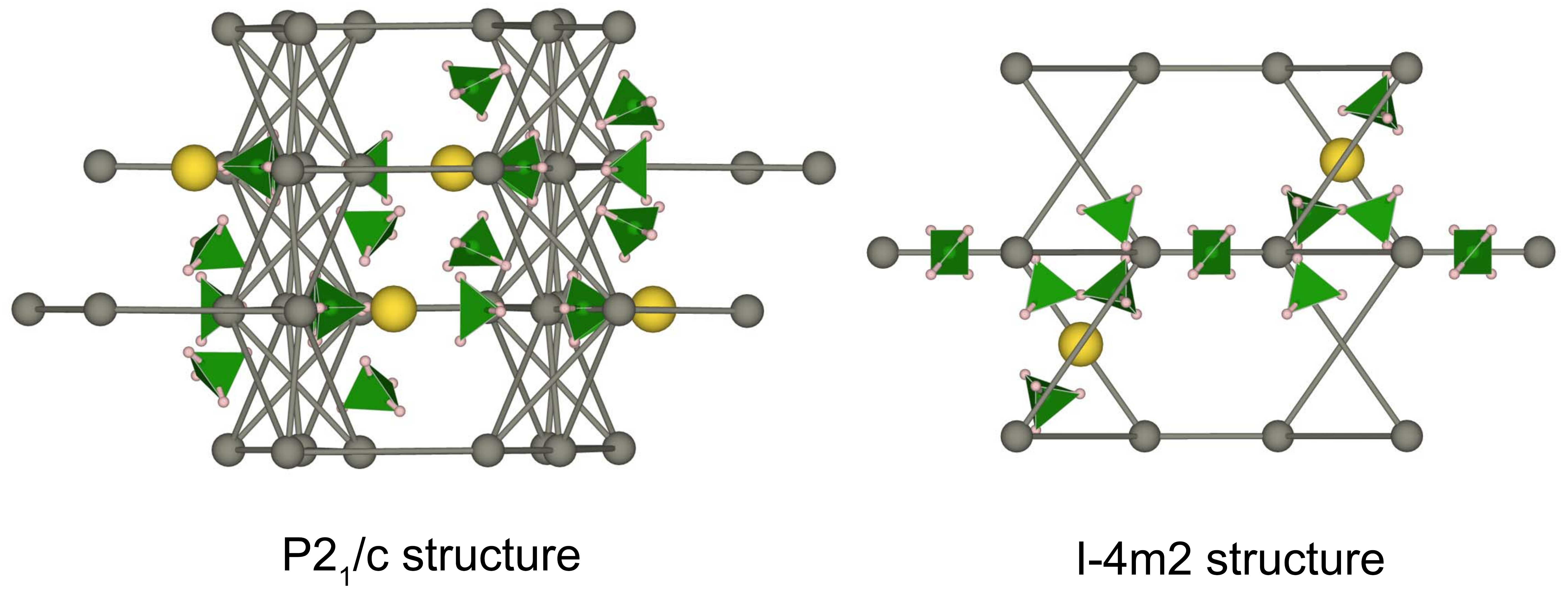}
  \caption{(Color online) Two low-energy phases of NaZn$_2$(BH$_4$)$_5$: $P2_1/c$ (left)
  and $I\overline{4}m2$ (right). Sodium and zinc and atoms are shown by yellow and gray
  spheres while [BH$_4$]$^-$ complex anions are shown by green tetrahedra. }\label{fig:NaZn2}
  \end{center}
\end{figure}

NaZn$_2$(BH$_4$)$_5$ was experimentally determined \cite{RavnsbaekLiNa_Zn} to crystallize in a $P2_1/c$ phase. The lowest-energy structure we discovered by the MHM runs for NaZn$_2$(BH$_4$)$_5$ belongs to the $I\overline{4}m2$ space group, which is thermodynamically less stable by just 0.2 kJ/mol cation than the $P2_1/c$ phase. The $I\overline{4}m2$ structure for NaZn$_2$(BH$_4$)$_5$ can also be obtained by relaxing the $I\overline{4}m2$ structure of LiZn$_2$(BH$_4$)$_5$ after replacing the Li cations by Na cations. Similar to the case of LiZn$_2$(BH$_4$)$_5$, the $I\overline{4}m2$ consists of one 3D framework while the $P2_1/c$ structure is composed of two inter-penetrated frameworks. \cite{RavnsbaekLiNa_Zn} The $I\overline{4}m2$ structure is therefore a low-density phase with the density $\rho = 0.54~{\rm g \times cm}^{-3}$, roughly one-half of the density  $\rho=1.15~{\rm g \times cm}^{-3}$ of the $P2_1/c$ structure. A geometrical illustration for the $P2_1/c$ and $I\overline{4}m2$ phases of NaZn$_2$(BH$_4$)$_2$ is given in Fig. \ref{fig:NaZn2}.

\begin{figure}[t]
  \begin{center}
    \includegraphics[width=8.25cm]{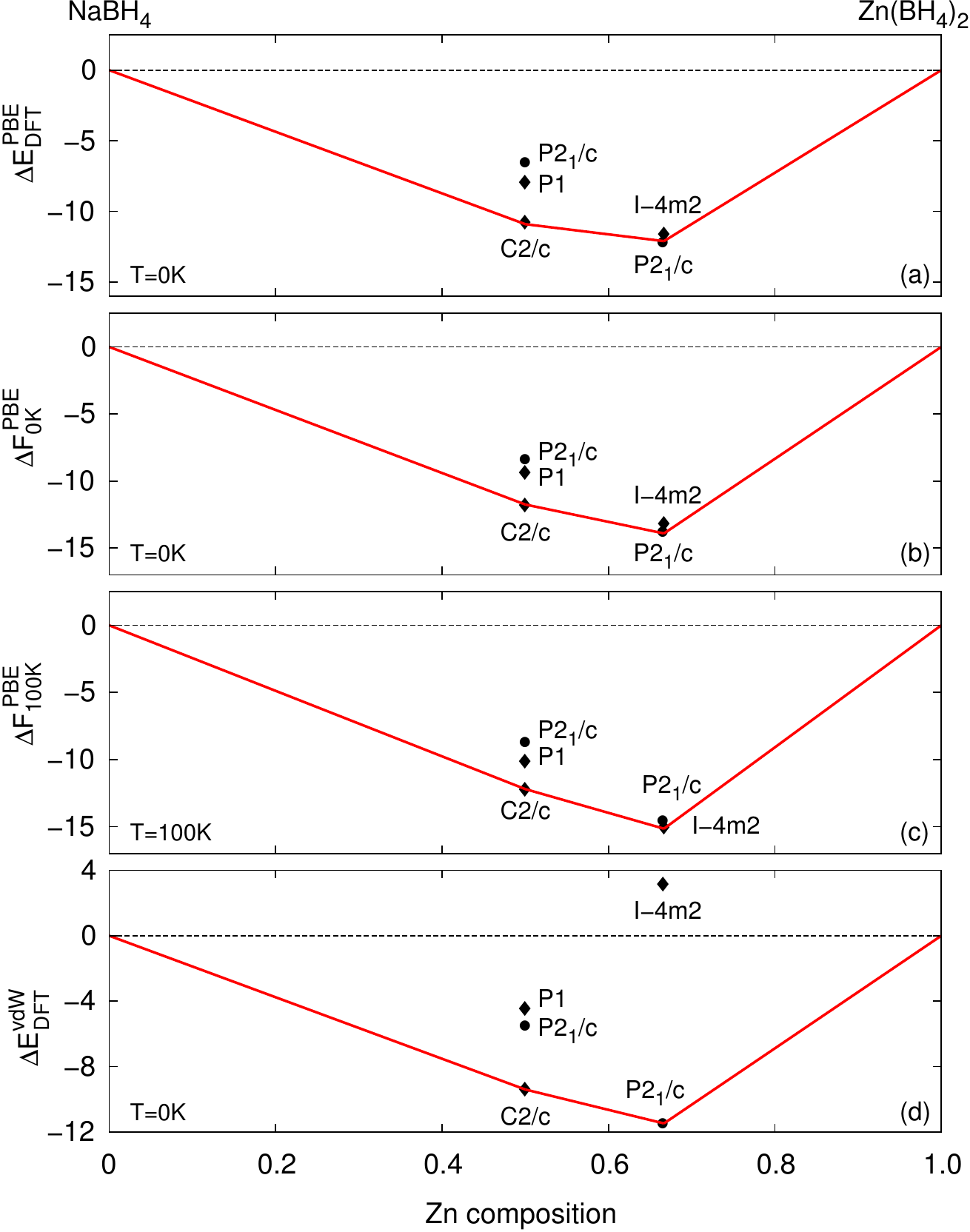}
  \caption{(Color online) Formation energies $\Delta E^{\rm PBE}_{\rm DFT}$,
  $\Delta F^{\rm PBE}_{\rm 0K}$, $\Delta F^{\rm PBE}_{\rm 100K}$, and
  $\Delta E^{\rm vdW}_{\rm DFT}$ (in units of kJ/mol cation) of Na/Zn
  double-cation borohydrides at 0K and 100K shown vs. Zn composition
  $x$. The structures studied are indicated by symbols with the space
  groups nearby. Diamonds/circles are used for
  theoretically-predicted/experimentally-reported phases. Constructed
  from the thermodynamically most stable structures, the red-solid
  lines form a convex hull, above which all structures are unstable. }\label{Fig:stableNaZn}
  \end{center}
\end{figure}

Based on the formation energies, the thermodynamic stability of the structures reported for both NaZn(BH$_4$)$_3$ and NaZn$_2$(BH$_4$)$_5$ was examined and presented in Fig. \ref{Fig:stableNaZn}. With the PBE functional, the $P2_1/c$ and $I\overline{4}m2$ phases of NaZn$_2$(BH$_4$)$_5$ are almost degenerate, being the two lowest-energy structures of this compound. For NaZn(BH$_4$)$_3$, $C2/c$ is the lowest-energy structure. Unlike Ref. \onlinecite{Aidhy:LiNaZn}, the formation energy of the $P2_1/c$ phase of NaZn(BH$_4$)$_3$ is always negative, indicating that this phase is stable and would not decompose into NaBH$_4$ and Zn(BH$_4$)$_2$. In addition, both the $P2_1/c$ and $P1$ structures were found to be unstable with respect to the decomposition into NaBH$_4$ and NaZn$_2$(BH$_4$)$_5$. The $C2/c$ phase, on the other hand, was found to be stable.

Similar to the case of Li/Zn borohydrides, the vdW interactions captured by vdW-DF2 calculations play an important role as it lifts the ``degeneracy" of the $P2_1/c$ and the $I\overline{4}m2$ structures of NaZn$_2$(BH$_4$)$_5$. In particular, the formation energy of the $I\overline{4}m2$ structure is slightly positive while $P2_1/c$ is the thermodynamically most stable structure with a strongly negative formation energy. To explain this behavior, we note that the $I\overline{4}m2$ structure consists of one 3D framework while the $P2_1/c$ structure is composed of two inter-penetrated frameworks that are not linked by bonds. The vdW interaction also reverts the energetic ordering of the $P2_1/c$ and the $P1$ structures of NaZn(BH$_4$)$_3$. Moreover, the $C2/c$ phase is again the lowest-energy structure of NaZn(BH$_4$)$_3$ and is stable with respect to the decomposition into NaBH$_4$ and NaZn$_2$(BH$_4$)$_5$. This result shows that from the theoretical point of view, the existence of the $C2/c$ phase of NaZn(BH$_4$)$_3$ is possible.

\subsection{Potassium/zinc borohydrides}

\begin{figure}[t]
  \begin{center}
    \includegraphics[width=8.25cm]{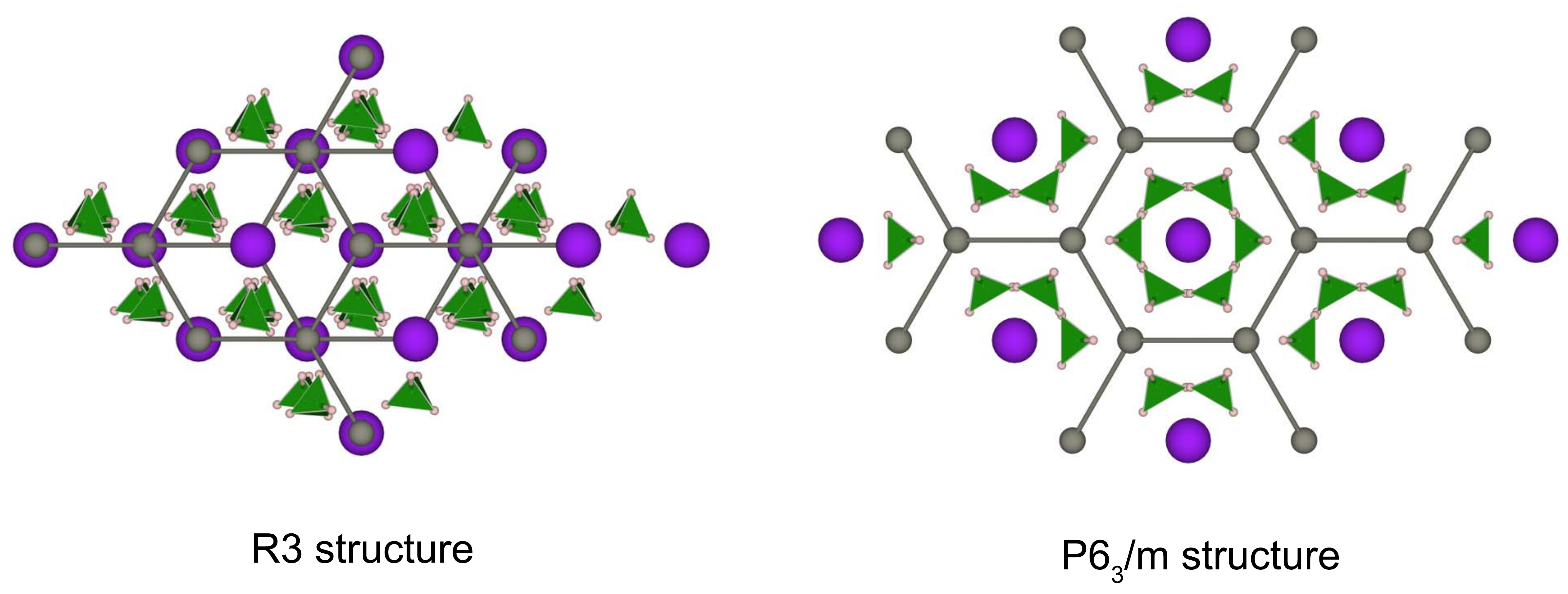}
  \caption{(Color online) Two low-temperature phases of KZn(BH$_4$)$_3$, the $R3$ phase
  (left) and $P6_3/m$ phase (right). Potassium and zinc atoms are shown by purple and gray
  spheres while green tetrahedra represent the [BH$_4$]$^-$ anions.}\label{fig:KZn}
  \end{center}
\end{figure}

\begin{figure}[b]
  \begin{center}
    \includegraphics[width = 5 cm]{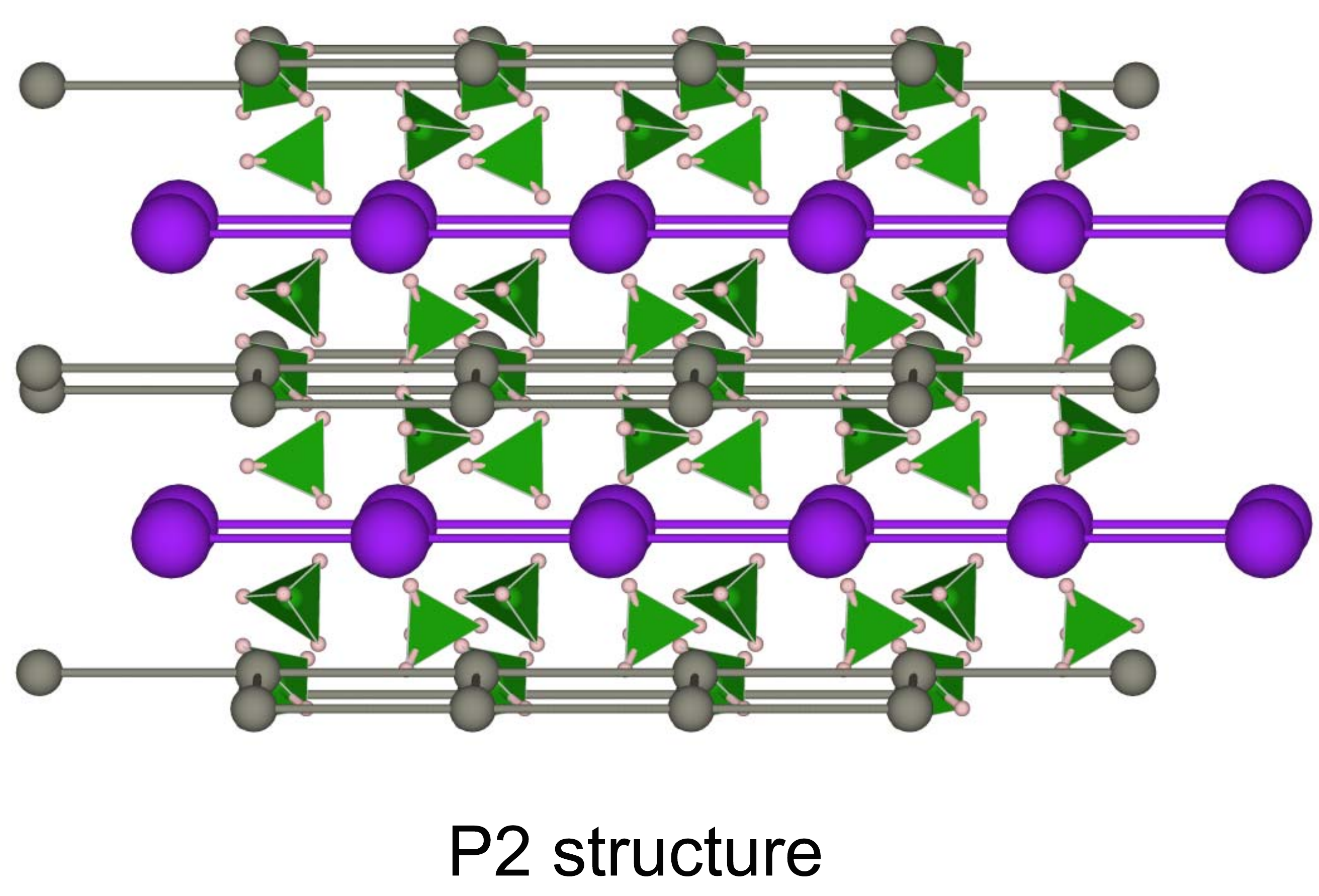}
  \caption{(Color online) The predicted $P2$ structure of KZn$_2$(BH$_4$)$_5$. Potassium
  and zinc atoms are shown by purple and gray spheres while green tetrahedra represent
  the [BH$_4$]$^-$ anions.}\label{fig:KZn2}
  \end{center}
\end{figure}

At 100 K, the experimentally synthesized crystalline KZn(BH$_4$)$_3$ sample was determined \cite{Cerny12_KZn} to be in a rhombohedral $R3$ phase. A related compound, KZn$_5$(BH$_4$)$_5$, on the other hand, has not been experimentally identified. Therefore, it would be interesting to explore, in a similar manner to the analysis performed for LiZn(BH$_4$)$_3$, if this hypothetical compound is stable or not. Our MHM runs for KZn(BH$_4$)$_3$ predicted a hexagonal $P6_3/m$ structure (no. 176) to be lowest in the DFT energy with the PBE functional $E^{\rm PBE}_{\rm DFT}$. Compared to the $R3$ structure, the $P6_3/m$ structure is thermodynamically more stable by $\simeq 3.5$ kJ/mol cation. The density of the $P6_3/m$ structure is $\rho=1.05~{\rm g \times cm}^{-3}$, slightly smaller than that of the $R3$ structure with $\rho=1.15~{\rm g \times cm}^{-3}$. For KZn$_2$(BH$_4$)$_5$, a monoclinic $P2$ structure (no. 3) with $\rho=1.09~{\rm g \times cm}^{-3}$ was predicted to be quite low in the DFT energy $E^{\rm PBE}_{\rm DFT}$ (see Table \ref{table_energy}). The geometries of the $P6_3/m$ and $R3$ structures for KZn(BH$_4$)$_3$ are shown in Fig. \ref{fig:KZn} while the geometry of the $P2$ structure of KZn$_2$(BH$_4$)$_5$ is shown in Fig. \ref{fig:KZn2}.

\begin{figure}[t]
  \begin{center}
    \includegraphics[width=8.25cm]{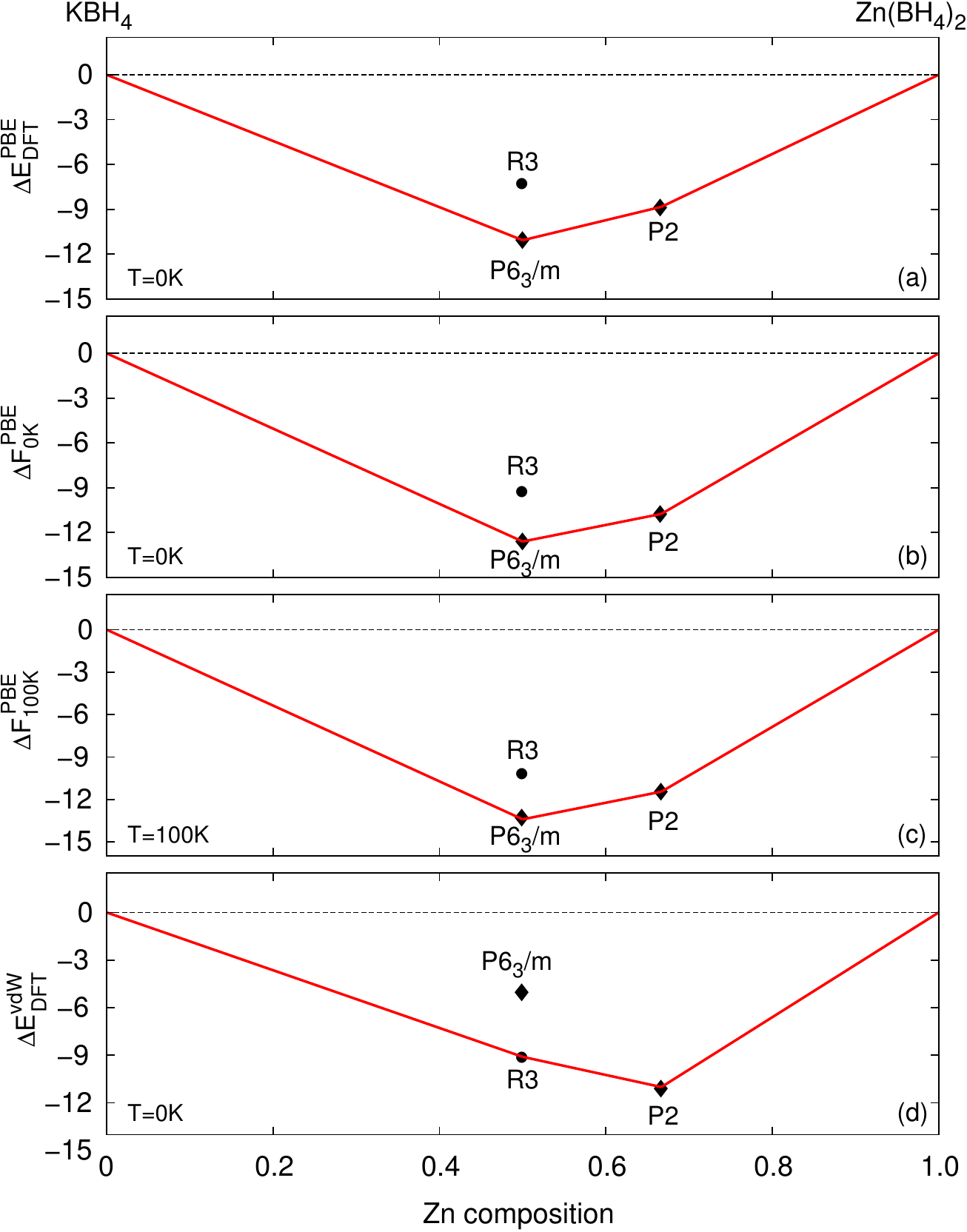}
  \caption{(Color online) Formation energies $\Delta E^{\rm PBE}_{\rm DFT}$,
  $\Delta F^{\rm PBE}_{\rm 0K}$, $\Delta F^{\rm PBE}_{\rm 100K}$, and
  $\Delta E^{\rm vdW}_{\rm DFT}$ (in units of kJ/mol cation) of K/Zn double-cation
  borohydrides at 0K and 100K shown vs. Zn composition $x$. The structures studied
  are indicated by symbols with the space
  groups nearby. Diamonds/circles are used for
  theoretically-predicted/experimentally-reported phases. Constructed
  from the thermodynamically most stable structures, the red-solid
  lines form a convex hull, above which all structures are unstable.}\label{Fig:stableKZn}
  \end{center}
\end{figure}

The thermodynamic stability of the $R3$ and $P6_3/m$ structures of KZn(BH$_4$)$_3$ and the $P2$ structure of KZn$_2$(BH$_4$)$_5$ was examined and the results are shown in Fig. \ref{Fig:stableKZn}. With the PBE functional, the energies of formation of these structures were determined to be all negative at 0K and 100K, implying that they are stable and would not decompose into KBH$_4$ and Zn(BH$_4$)$_2$ [see Figs. \ref{Fig:stableKZn}(a)-(c)]. Both the $R3$ and the $P6_3/m$ structures of  KZn(BH$_4$)$_3$ are located below the line connecting KBH$_4$ and the $P2$ phase of KZn$_2$(BH$_4$)$_5$, thus they are also stable with respect to the decomposition into KBH$_4$ and KZn$_2$(BH$_4$)$_5$. Similarly, the $P2$ phase of KZn$_2$(BH$_4$)$_5$ is stable and would not decompose into KZn(BH$_4$)$_3$ and Zn(BH$_4$)$_2$. Fig. \ref{Fig:stableKZn}(d) indicates that the inclusion of the vdW interactions reverts again the energetic ordering of the $R3$ and the $P6_3/m$ structures of  KZn(BH$_4$)$_3$. Consequently, the experimentally reported $R3$ structure is found to be the thermodynamically most stable phase of KZn(BH$_4$)$_3$. For KZn$_2$(BH$_4$)$_5$, the formation energy of the $P2$ structure remains strongly negative, indicating that this phase is thermodynamically stable. The existence of this  hypothetical compound KZn$_2$(BH$_4$)$_5$ is therefore theoretically possible.

\section{Conclusions}
We performed a systematic study on the  thermodynamic stability of a series of Li/Zn, Na/Zn, and K/Zn double-cation borohydrides at low temperatures. Some of them were recently synthesized, others are hypothetical. Using the minima-hopping method, we discovered two new lowest-energy structures for NaZn(BH$_4$)$_3$, the compound which was experimentally synthesized and KZn$_2$(BH$_4$)$_5$, a hypothetical compound. The lowest-energy structures of these compounds, which belong to the $C2/c$  and $P2$ space groups, respectively, are stable with respect to the decomposition into the related compounds. For NaZn(BH$_4$)$_3$ which was experimentally synthesized, the possible existence of the $C2/c$ phase can change the temperature and the pressure at which the decomposition reaction occur. Given that such a reaction is identified, calculations for these changes can readily be performed by, for example, the method described in Ref. \onlinecite{Alapati07b}. For the hypothetical LiZn(BH$_4$)$_3$ compound, we reach the same conclusion of a previous study \cite{Aidhy:LiNaZn} that the theoretically proposed $P1$ structure is lowest in energy. We also found that this structure is unstable and would decompose into LiBH$_4$ and LiZn$_2$(BH$_4$)$_5$. Therefore, this instability can be viewed as a supporting evidence for the hypothesis that LiZn(BH$_4$)$_3$ can not be formed. On the other hand, the experimentally-reported structures of LiZn$_2$(BH$_4$)$_5$, NaZn$_2$(BH$_4$)$_5$, and KZn(BH$_4$)$_3$ are also predicted to be thermodynamically most stable structures.

While lattice vibrations were found to play a minor role in the energetic ordering of the studied structures, van der Waals interactions, on the other hand, seem to be critically important. We found that the PBE functional is not sufficient to capture the non-bonding interactions between the constituent 3D frameworks of the complex structures of LiZn$_2$(BH$_4$)$_5$ and NaZn$_2$(BH$_4$)$_5$. Consequently, the energetic ordering of the examined low-energy structures determined at the PBE level is changed. To overcome this problem, we suggest that vdW-DF2, a recently developed non-local density functional, is suitable to access the stability of the low-energy structures of these double-cation borohydrides.

\begin{acknowledgments}
The authors thank Radovan \v{C}ern\'{y} for valuable expert discussions. Useful discussions and suggestions from Dipulneet Aidhy, Nguyen-Manh Duc, Atsushi Togo, and Aleksey Kolmogorov are appreciated. T. D. H, M. A., and S. G. gratefully acknowledge the financial support from the Swiss National Science Foundation. Work by VNT is supported by the Vietnamese NAFOSTED program No. 103.02-2011.20. L. M. W. gratefully acknowledges financial support from the Department of Energy under Contract No. DE-FG02-06ER46297. The computational work was performed at the Swiss National Supercomputing Center, Hanoi University of Science and Technology, the Research Computing at the University of South Florida, and \'{E}cole Polytechnique F\'{e}d\'{e}rale de Lausanne.
\end{acknowledgments}


\end{document}


\title{Supplemental Material: Thermodynamic stability of alkali metal/zinc double-cation borohydrides at low temperatures}
\author{Tran Doan Huan}
\affiliation{Department of Physics, Universit\"{a}t Basel, Klingelbergstrasse 82, 4056 Basel, Switzerland}
\author{Maximilian Amsler}
\affiliation{Department of Physics, Universit\"{a}t Basel, Klingelbergstrasse 82, 4056 Basel, Switzerland}
\author{Riccardo Sabatini}
\affiliation{Theory and Simulation of Materials, \'{E}cole Polytechnique F\'{e}d\'{e}rale de Lausanne, Station 12, 1015 Lausanne, Switzerland}
\author{Vu Ngoc Tuoc}
\affiliation{Institute of Engineering Physics, Hanoi University of Science and Technology, 1 Dai Co Viet Road, Hanoi, Vietnam}
\author{Nam Ba Le}
\affiliation{Department of Physics, University of South Florida, 4202 E. Fowler Ave., Tampa, FL 33620, USA}
\author{Lilia M. Woods}
\affiliation{Department of Physics, University of South Florida, 4202 E. Fowler Ave., Tampa, FL 33620, USA}
\author{Nicola Marzari}
\affiliation{Theory and Simulation of Materials, \'{E}cole Polytechnique F\'{e}d\'{e}rale de Lausanne, Station 12, 1015 Lausanne, Switzerland}
\author{Stefan Goedecker}
\email{stefan.goedecker@unibas.ch}
\affiliation{Department of Physics, Universit\"{a}t Basel, Klingelbergstrasse 82, 4056 Basel, Switzerland}

\date{\today}

\maketitle

\begin{table*}[t]
\begin{center}
\caption{Volume change $\Delta V$, given in percents, of some experimentally
reported compounds optimized  with various exchange-correlation functionals.}
\label{table_XC}
\begin{tabular}{llrrrrrr}
\hline
\hline
Compounds              &Space group & PBE    & PBEsol & LDA      & vdW-DF2 & DFT-D2  & DFT-TS \\
\hline
LiZn$_2$(BH$_4$)$_5$   & $Cmca$     & 2.0    & $-4.1$ & $-12.1$  & $0.5$   & $-8.6$ &  $-10.0$ \\
NaZn(BH$_4$)$_3$       & $P2_1/c$   & 14.3   & $0.4$  & $-11.8$  & $-1.2$  & $-18.9$ & $-16.2$\\
NaZn$_2$(BH$_4$)$_5$   & $P2_1/c$   & 6.4    & $-0.7$  & $-15.4$ & $-1.7$  & $-11.4$  & $-22.7$\\
KZn(BH$_4$)$_3$        & $R3$       & 16.5   & $5.7$  & $-7.5$   & $8.1$   & $-3.8$ & $-7.8$ \\
LiBH$_4$               & $Pnma$     & $-2.4$ & $-6.5$ & $-14.0$  & $-3.7$  & $-15.7$ & $-17.1$\\
NaBH$_4$               & $P4_2/nmc$ & 0.6    & $-4.8$ & $-11.6$  & $ 0.6$  & $-16.2$ & $-25.2$\\
KBH$_4$                & $P4_2/nmc$ & 3.8    & $-2.3$ & $-10.6$  & $0.9 $  & $-7.0$  & $-35.0$\\
\hline
\hline
\end{tabular}
\end{center}
\end{table*}

\begin{table*}[t]
\begin{center}
\caption{Geometrical parameters (lattice parameters $a$, $b$, and $c$ in Angstrom, angle $\beta$
in degree, and volume difference $\Delta V$ in percents) obtained by computationally optimizing
the experimentally-reported low-energy structures of the examined compounds with PBEsol XC
functional and vdW-DF2 density functional. Calculations were performed by {\sc vasp}.} \label{table_validate1}
\begin{tabular}{l l r r r r r r r r r r r r r r r}
\hline
\hline
\multirow{2}{*}{Compound} & \multirow{2}{*}{Space group} & & \multicolumn{4}{c}{PBESol} &&\multicolumn{4}{c}{vdW-DF2} & & \multicolumn{4}{c}{Experiments}\\
\cline{4-7}\cline{9-12}\cline{14-17}
& & & $a$ & $b$ &$c$ & $\beta$ & & $a$ & $b$& $c$& $\beta$ & &$a$ & $b$ &$c$ &$\beta$\\
\hline
LiZn$_2$(BH$_4$)$_5$ &  $Cmca$  & & 8.45 & 17.69 & 15.12 & 90     & & 8.71 & 17.50 & 15.67 & 90    & & 8.59 & 17.86 & 15.35 & 90 \\
NaZn(BH$_4$)$_3$ & $P2_1/c$     & & 7.13 & 5.58  & 17.65 & 100.52 & & 8.83 & 4.62  & 16.89 & 99.8  & & 8.27 & 4.52  & 18.76 & 101.7 \\
NaZn$_2$(BH$_4$)$_5$ & $P2_1/c$ & & 9.56 & 16.22 & 8.97  & 109.75 & & 8.69 & 17.12 & 9.34  & 111.5 & & 9.40 & 16.64 & 9.14  & 112.7 \\
KZn(BH$_4$)$_3$ &  $R3$         & & 7.55 & 7.55  & 11.86 & 90     & & 7.65 & 7.65  & 11.80 & 90    & & 7.63 & 7.63  & 10.98 & 90 \\
LiBH$_4$ &   $Pnma$             & & 7.22 & 4.31  & 6.51  & 90     & & 6.90 & 4.50  & 6.72  & 90    & & 7.18 & 4.44  & 6.80  & 90 \\
NaBH$_4$ &   $P4_2/nmc$         & & 4.26 & 4.26  & 5.77  & 90     & & 4.35 & 4.35  & 5.85  & 90    & & 4.33 & 4.33  & 5.87  & 90 \\
KBH$_4$ &    $P4_2/nmc$         & & 4.64 & 4.64  & 6.52  & 90     & & 4.72 & 4.72  & 6.60  & 90    & & 4.70 & 4.70  & 6.60  & 90 \\
\hline
\hline
\end{tabular}
\end{center}
\end{table*}

\begin{table*}[t]
\begin{center}
\caption{Geometrical parameters (lattice parameters $a$, $b$, and $c$ in Angstrom, angle $\beta$ in degree, and volume difference $\Delta V$ in
percents) obtained by computationally optimizing the experimentally-reported low-energy structures of the examined compounds with {\sc vasp} and {\sc quantum espresso}. Calculations were performed with vdW-DF2.} \label{table_validate2}
\begin{tabular}{llrrrrrrrrrrrrrrrr}
\hline
\hline
\multirow{2}{*}{Compound}&\multirow{2}{*}{Space group}&&\multicolumn{5}{c}{Calculations with VASP}&\multicolumn{5}{c}{Calculations with QE} & & \multicolumn{4}{c}{Experiments}\\
\cline{3-7}\cline{9-13}\cline{15-18}
&& $a$& $b$ & $c$ & $\beta$  &$\Delta V$& &$a$  & $b$ &$c$ & $\beta$ &$\Delta V$ &  &$a$ &$b$ &$c$ &$\beta$ \\
\hline
LiZn$_2$(BH$_4$)$_5$ &  $Cmca$    & 8.71 & 17.50 & 15.67 & 90    & 0.5    & & 8.76 & 17.62 & 15.78 & 90    & 2.4 && 8.59 & 17.86 & 15.35 & 90 \\
NaZn(BH$_4$)$_3$ & $P2_1/c$       & 8.83 & 4.62  & 16.89 & 99.8  & $-1.2$ & & 8.84 & 4.65  & 17.14 & 100.0 & 1.1 && 8.27 & 4.52 & 18.76 & 101.7 \\
NaZn$_2$(BH$_4$)$_5$  & $P2_1/c$  & 8.69 & 17.12 & 9.34  & 111.5 & $-1.7$ & & 9.07 & 17.05 & 9.36  & 112.2 & 1.7 && 9.40 & 16.64 & 9.14 & 112.7 \\
KZn(BH$_4$)$_3$ &  $R3$           & 7.65 & 7.65  & 11.80 & 90    & 8.1    & & 7.78 & 7.78  & 11.88 &  90   & 12.6&& 7.63  & 7.63  & 10.98 & 90 \\
LiBH$_4$ &   $Pnma$               & 6.90 & 4.50  & 6.72  & 90    & $-3.7$ & & 6.96 & 4.54  & 6.72  & 90    & $-2.1$&& 7.18  & 4.44  & 6.80 & 90 \\
NaBH$_4$ &   $P4_2/nmc$           & 4.35 & 4.35  & 5.85  & 90    & 0.6    & & 4.36 & 4.36  & 5.86  & 90    &1.2&& 4.33 & 4.33 & 5.87 & 90 \\
KBH$_4$ &    $P4_2/nmc$           & 4.72 & 4.72  & 6.60  & 90    & 0.9 & & 4.72 & 4.72  &6.60   & 90    &$-3.4$&& 4.70 & 4.70 & 6.60 & 90 \\
\hline
\hline
\end{tabular}
\end{center}
\end{table*}

\begin{table}[H]
\begin{center}
\caption{$I\overline{4}m2$ structure of LiZn$_2$(BH$_4$)$_5$ with $a=b=8.396$ \AA, $c=16.926$ \AA, $\alpha=\beta=\gamma=90^\circ$. } \label{Table_SG119LiZn}
\begin{tabular}{llccc}
\hline
\hline
Atom & WP  & $x$ & $y$ & $z$ \\
\hline
Li & 2d &  0.00000 & 0.50000 & 0.75000  \\
Zn & 4e &  0.00000 & 0.00000 & 0.13400  \\
B & 8i  & -0.23449 & 0.00000 & 0.18952  \\
B & 2a  &  0.00000 & 0.00000 & 0.00000  \\
H & 16j &  0.37333 & 0.33797 & 0.32746  \\
H & 8i  &  0.35451 & 0.00000 & 0.14971  \\
H & 8i  &  0.12618 & 0.00000 & -0.03731 \\
H & 8i  & -0.24850 & 0.00000 & 0.26071  \\
\hline
\hline
\end{tabular}
\end{center}
\end{table}

\begin{table}[H]
\begin{center}
\caption{$C2/c$ structure of NaZn(BH$_4$)$_3$ with $a=7.793$ \AA, $b=15.723$ \AA, $c=7.680$ \AA, $\alpha=\gamma=90^\circ$, $\beta=110.418^\circ$. } \label{Table_SG15NaZn}
\begin{tabular}{llccc}
\hline
\hline
Atom & WP  & $x$ & $y$ & $z$ \\
\hline
Na & 4e& 0.00000& -0.05711 &0.25000 \\
Zn & 4e& 0.00000& -0.37109 & 0.25000 \\
B & 8f &-0.24773& 0.04904 &0.33521 \\
B & 4e& 0.00000 &-0.23185 &0.25000 \\
H & 8f&  0.23169& 0.47318 &0.30728 \\
H & 8f& -0.09158& 0.07548 &0.40084 \\
H & 8f& -0.31184& 0.08602 &0.18126 \\
H & 8f& -0.32512& 0.06702 &0.44784 \\
H & 8f& -0.36324& 0.30786 &0.26099 \\
H & 8f& -0.46155& 0.22504 &0.39673 \\
\hline
\hline
\end{tabular}
\end{center}
\end{table}

\begin{table}[H]
\begin{center}
\caption{$I\overline{4}m2$ structure of NaZn$_2$(BH$_4$)$_5$ with $a=b=8.863$ \AA, $c=17.823$ \AA, $\alpha=\beta=\gamma=90^\circ$.} \label{Table_SG119NaZn2}
\begin{tabular}{llccc}
\hline
\hline
Atom & WP  & $x$ & $y$ & $z$ \\
\hline
Na & 2d& 0.00000  &  0.50000 & 0.75000  \\
Zn & 4e &0.00000  &  0.00000 & 0.12723 \\
B & 8i &-0.22178  &  0.00000 & 0.18054 \\
B & 2a &0.00000   &  0.00000 & 0.00000 \\
H & 16j& -0.38010 & -0.34645 & 0.33640 \\
H & 8i &0.33654   &  0.00000 & 0.14357 \\
H & 8i & 0.11954  &  0.00000 & -0.03543 \\
H & 8i &-0.23158  &  0.00000 & 0.24830 \\
\hline
\hline
\end{tabular}
\end{center}
\end{table}

\begin{table}[H]
\begin{center}
\caption{$P6_3/m$ structure of KZn(BH$_4$)$_3$ with $a=b=8.361$ \AA, $c=7.820$ \AA, $\alpha=\beta=90^\circ$, $\gamma=120^\circ$.}\label{Table_SG176KZn}
\begin{tabular}{llccc}
\hline
\hline
Atom & WP  & $x$ & $y$ & $z$ \\
\hline
K  & 2b & 0.00000  &0.00000   & 0.00000  \\
Zn & 2c & 0.33333  & 0.66667  & 0.25000 \\
B  & 6h & 0.18024  & 0.36048   & 0.25000  \\
H  & 12i& -0.22693 & -0.45387 & -0.11496  \\
H  & 6h & -0.27308 &-0.01277  & 0.25000  \\
H  & 6h & -0.27307 & -0.26031  & 0.25000  \\
\hline
\hline
\end{tabular}
\end{center}
\end{table}

\begin{table}[H]
\begin{center}
\caption{$P2$ structure of KZn$_2$(BH$_4$)$_5$ with $a=7.146$ \AA, $b=5.130$ \AA, $c=10.151$ \AA, $\alpha = \gamma=90^\circ$, $\beta=88.741^\circ$.}\label{Table_SG03KZn2}
\begin{tabular}{llccc}
\hline
\hline
Atom & WP  & $x$ & $y$ & $z$ \\
\hline
K  & 1c &0.50000  &-0.28780 &0.00000 \\
Zn & 2e &-0.04620 &0.09929  &-0.28148 \\
B  & 2e &0.29681  &0.25477  &0.17862  \\
B  & 1b &0.00000  &0.16741  & 0.50000  \\
B  & 2e &0.16888  &-0.14806  & -0.19612  \\
H  & 2e &0.31447  &0.22295   &0.06063 \\
H  & 2e &0.12578  &0.02371   &0.45644 \\
H  & 2e &0.14255  &-0.21993  & -0.08374  \\
H  & 2e &0.41437  &0.38313   &0.23277 \\
H  & 2e &-0.07793 & 0.31402  & 0.42239  \\
H  & 2e &-0.14432 & 0.37736  & -0.19422  \\
H  & 2e &0.17621  & 0.09776  &-0.19369  \\
H  & 2e &0.29818  & 0.03408  &0.23206  \\
H  & 2e &0.30971  & -0.21894 & -0.25384 \\
H  & 2e &0.03511  & -0.23848 & -0.26096 \\
\hline
\hline
\end{tabular}
\end{center}
\end{table}

\begin{figure}[ht]
  \begin{center}
    \includegraphics[width=11cm]{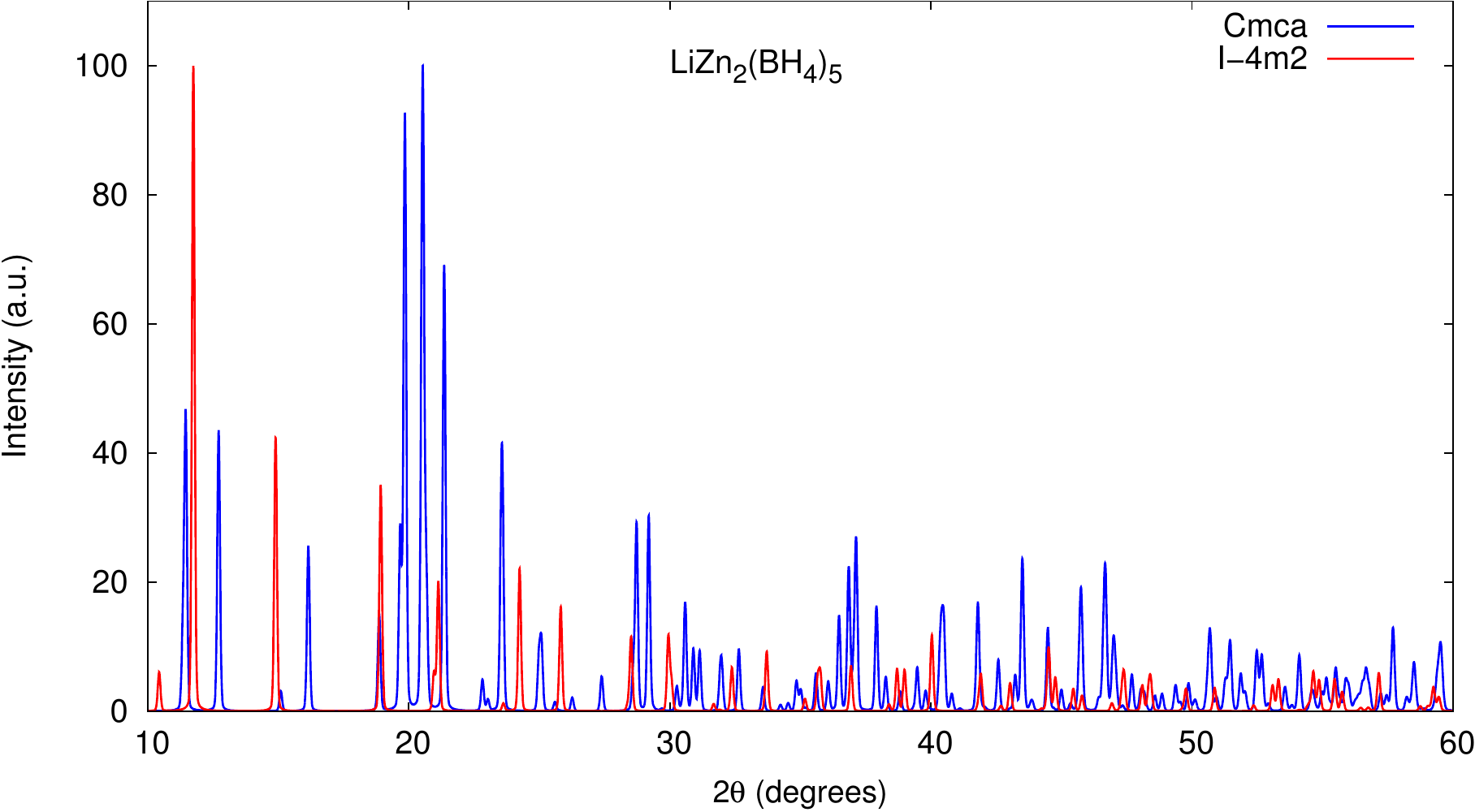}
  \caption{(Color online) Simulated XRD patterns of the $Cmca$ and $I\overline{4}m2$ phases of LiZn$_2$(BH$_4$)$_5$.}
  \end{center}
\end{figure}

\begin{figure}[ht]
  \begin{center}
    \includegraphics[width=11cm]{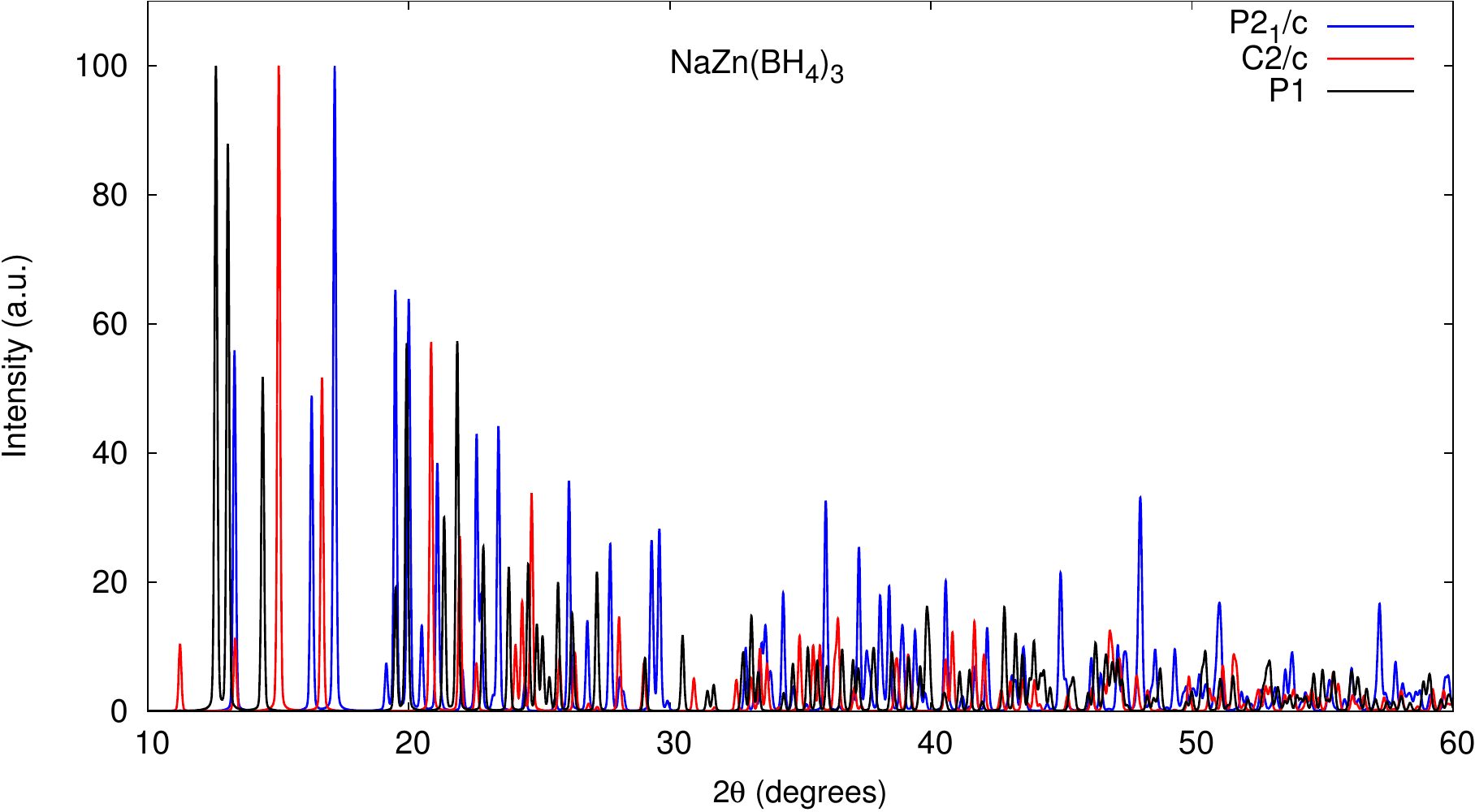}
  \caption{(Color online) Simulated XRD patterns of the $P2_1/c$, $C2/c$, and $P1$ phases of NaZn(BH$_4$)$_3$.}
  \end{center}
\end{figure}

\begin{figure}[ht]
  \begin{center}
    \includegraphics[width=11cm]{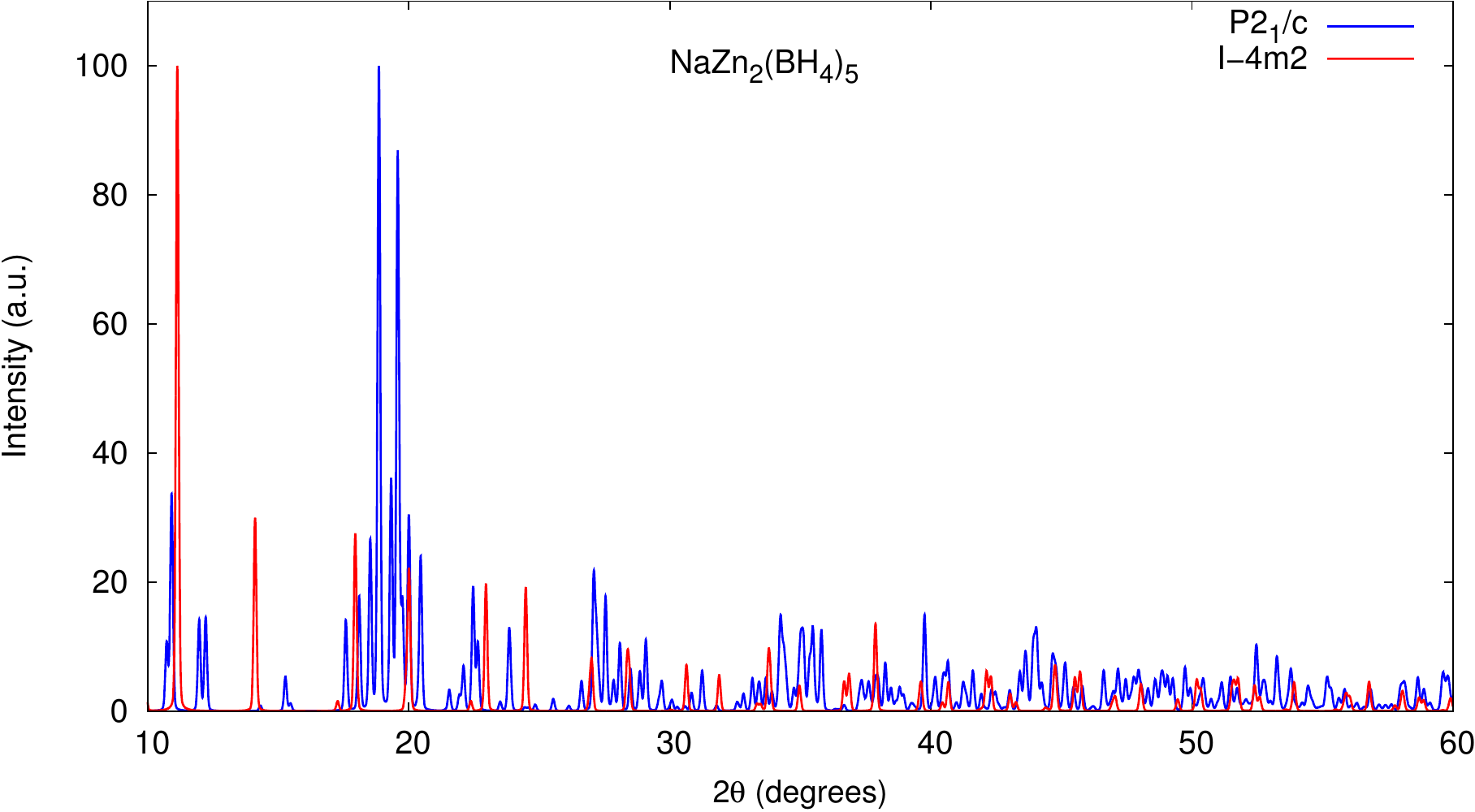}
  \caption{(Color online) Simulated XRD patterns of the $P2_1/c$ and $I\overline{4}m2$ phases of NaZn$_2$(BH$_4$)$_5$.}
  \end{center}
\end{figure}

\begin{figure}[ht]
  \begin{center}
    \includegraphics[width=11cm]{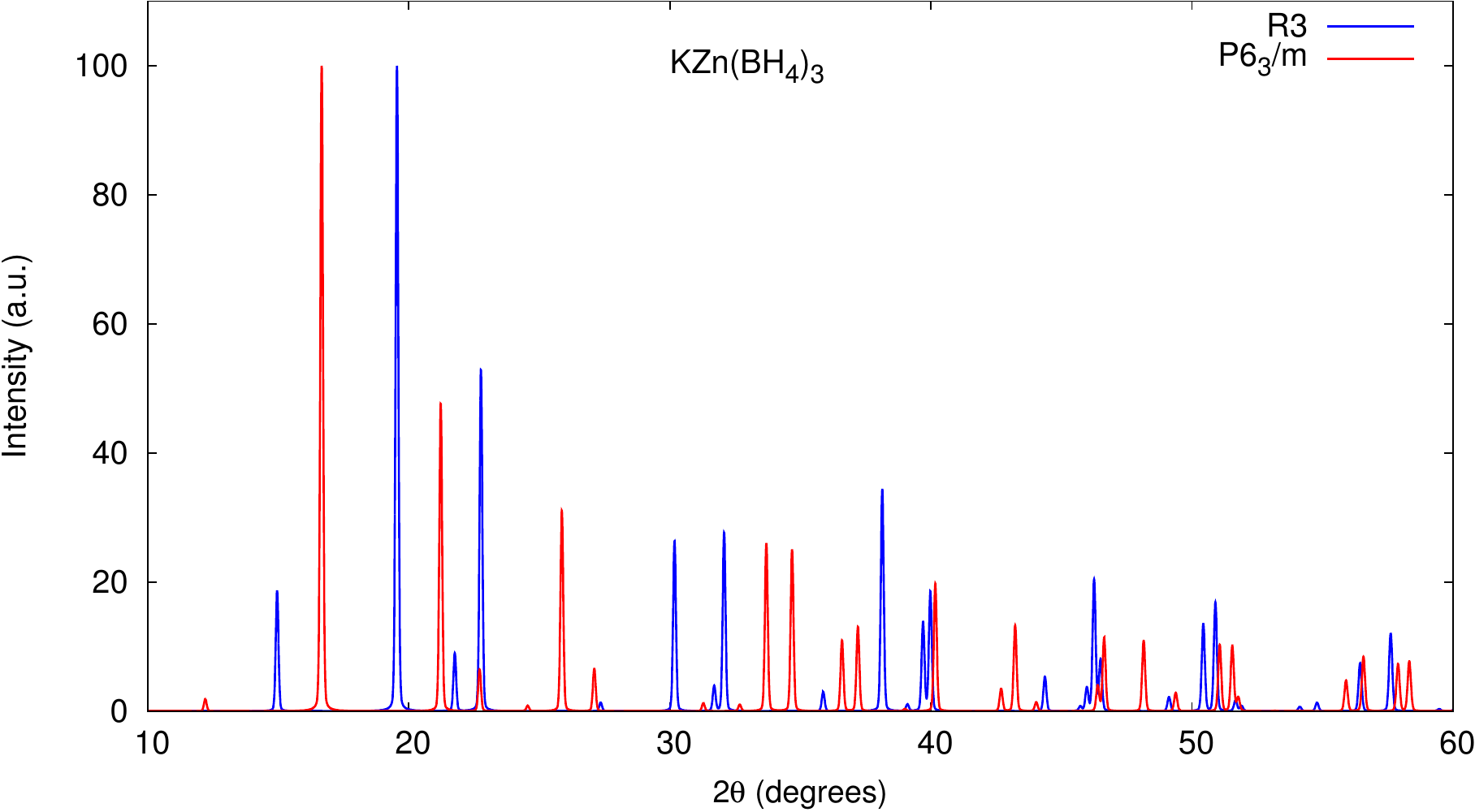}
  \caption{(Color online) Simulated XRD patterns of the $R3$ and $P6_3/m$ phases of KZn(BH$_4$)$_3$.}
  \end{center}
\end{figure}

\begin{figure}[ht]
  \begin{center}
    \includegraphics[width=11cm]{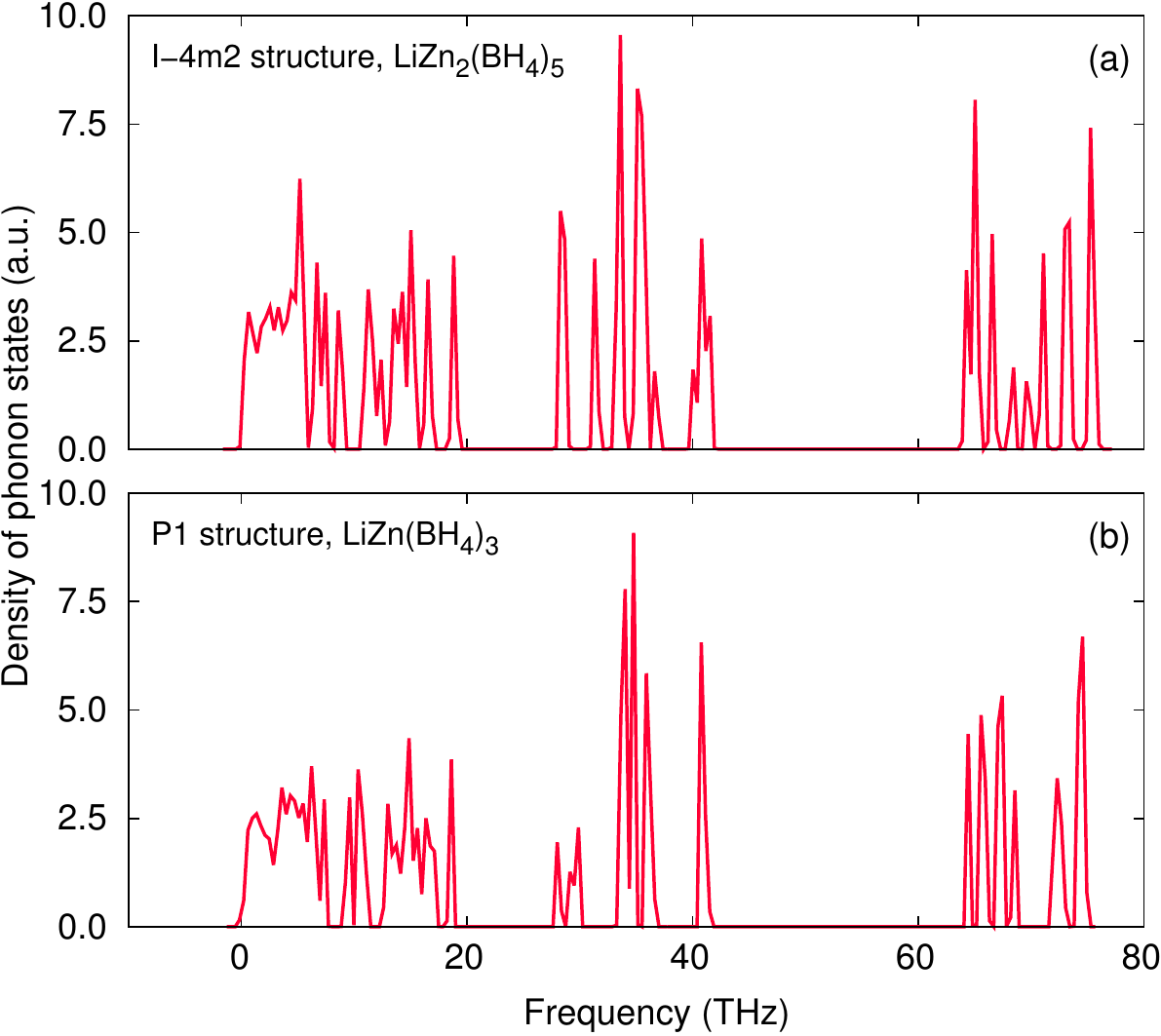}
  \caption{(Color online) Density of phonon states of (a) the $I\overline{4}m2$ structure of LiZn$_2$(BH$_4$)$_5$ and (b) the $P1$ structure of LiZn(BH$_4$)$_3$.}\label{fig:phdos_LiZn}
  \end{center}
\end{figure}

\begin{figure}[ht]
  \begin{center}
   \includegraphics[width=11cm]{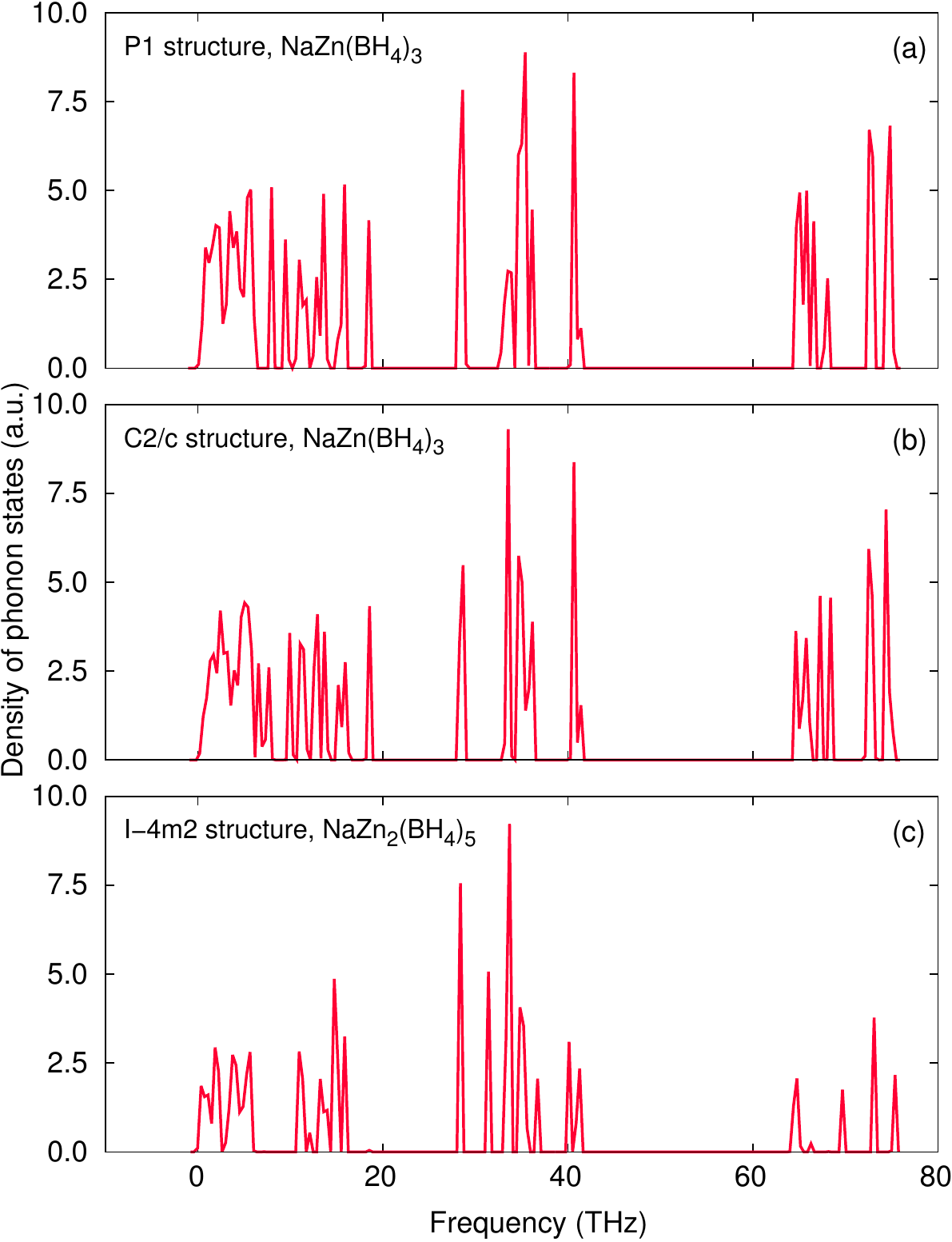}
  \caption{(Color online) Density of phonon states of (a) the $P1$ structure of NaZn(BH$_4$)$_3$, (b) the $C2/c$ structure of
  NaZn(BH$_4$)$_3$ and (b) the $I\overline{4}m2$ structure of NaZn$_2$(BH$_4$)$_5$.}\label{fig:phdos_NaZn}
  \end{center}
\end{figure}

\begin{figure}[ht]
  \begin{center}
    \includegraphics[width=11cm]{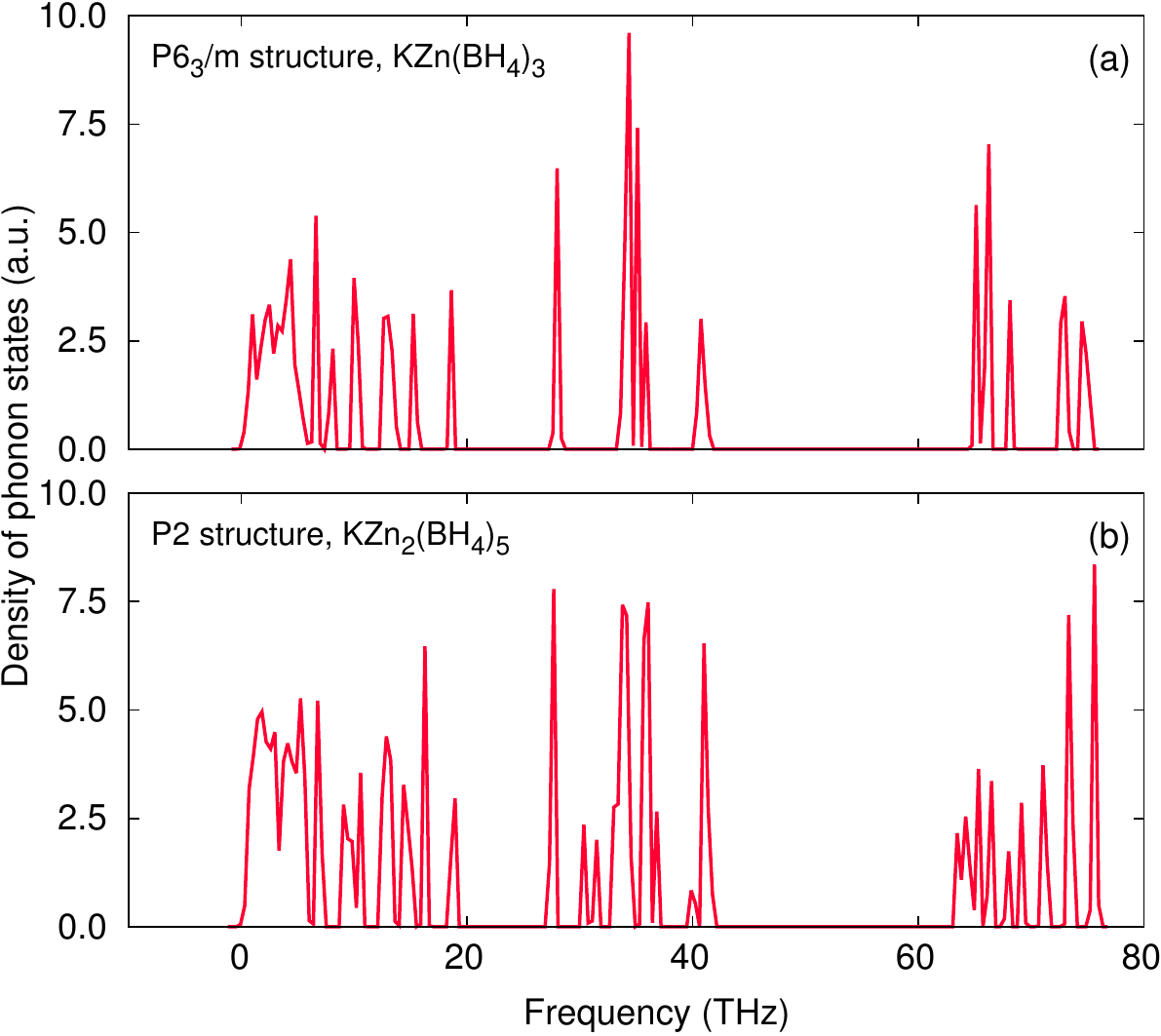}
  \caption{(Color online) Density of phonon states of (a) the $P6_3/m$ structure of
  KZn(BH$_4$)$_3$ and (b) the $P2$ structure of KZn$_2$(BH$_4$)$_5$.}\label{fig:phdos_KZn}
  \end{center}
\end{figure}